\documentclass[longauth,desactivate]{aa}

\usepackage{graphicx,upgreek}
\usepackage[varg]{txfonts}
\usepackage[dvipsnames]{xcolor}
\usepackage[colorlinks=true,citecolor=blue, urlcolor=blue, linkcolor=blue]{hyperref}
\usepackage{ulem}  

\makeatletter
\renewcommand*\aa@pageof{, page \thepage{} of \pageref*{LastPage}}
\makeatother

\newcommand{\RHill}{\ensuremath{{R_{\textnormal{Hill}}}}\xspace}
\newcommand{\kB}{\ensuremath{k_{\textnormal{B}}}\xspace}
\newcommand{\Hm}{\ensuremath{\rm H ^{-}}\xspace}

\begin{document}

   \title{Multi-frequency observations of PDS\,70c: Radio emission mechanisms in the circumplanetary environment}

   \titlerunning{Multi-frequency observations of PDS\,70c}
   
   \authorrunning{Dom\'inguez-Jamett et al.}

   \author{Oriana~Dom\'inguez-Jamett\inst{\ref{UCH}}
          \and
          Simon~Casassus\inst{\ref{UCH},\ref{DO}}
          \and
          Hauyu~Baobab~Liu\inst{\ref{Baobab1},\ref{Baobab2}}
          \and
          Yuhiko~Aoyama\inst{\ref{yuhiko1}}
          \and
          Miguel~C\'arcamo\inst{\ref{USACH},\ref{DO}}
          \and
          Philipp~Weber\inst{\ref{USACH3},\ref{yems},\ref{USACH2}}
          \and
          Ond\v{r}ej~Chrenko\inst{\ref{charles}}
          \and
          Gabriel-Dominique~Marleau\inst{\ref{Bern},\ref{MPIA},\ref{UDE}}
          \and
          Barbara~Ercolano\inst{\ref{LMU},\ref{garching}}
          \and
          Judit~Szul\'agyi\inst{\ref{zuerich},\ref{ETH}}
          }

          \institute{Departamento de Astronom\'{\i}a, Universidad de Chile, Casilla 36-D, Santiago, Chile\\
            \email{oriana.dominguez@ug.uchile.cl} \label{UCH}
            \and
            {Data Observatory Foundation, Eliodoro Y\'a\~{n}ez 2990, Providencia, Santiago, Chile} \label{DO}
            \and
            {Department of Physics, National Sun Yat-Sen University, No.~70, Lien-Hai Road, Kaohsiung City 80424, Taiwan, R.O.C.} \label{Baobab1}
            \and 
            {Center of Astronomy and Gravitation, National Taiwan Normal University, Taipei 116, Taiwan} \label{Baobab2}
            \and
            {School of Physics and Astronomy, Sun Yat-sen University, Guangdong 519082, People's Republic of China} \label{yuhiko1}
            \and
            University of Santiago of Chile (USACH), Faculty of Engineering, Computer Engineering Department, Chile \label{USACH}
            \and
            {Departamento de Física, Universidad de Santiago de Chile, Av. V\'ictor Jara 3493, Santiago, Chile} \label{USACH3}
            \and
            {Millennium Nucleus on Young Exoplanets and their Moons (YEMS), Chile} \label{yems}
            \and
            {Center for Interdisciplinary Research in Astrophysics and Space Exploration (CIRAS), Universidad de Santiago de Chile, Chile}  \label{USACH2}
            \and
            Charles University, Faculty of Math and Physics, Astronomical Institute, V Hole\v{s}ovi\v{c}k\'{a}ch 747/2, 180 00 Prague 8, Czech Republic\label{charles}  
            \and
           Division of Space Research \&\ Planetary Sciences, Physics Institute, University of Bern, Gesellschaftsstr.~6, 3012 Bern, Switzerland\label{Bern}  
           \and
           {Max-Planck-Institut f\"ur Astronomie, K\"onigstuhl 17, 69117 Heidelberg, Germany} \label{MPIA}
            \and
            {Fakult\"at f\"ur Physik, Universit\"at Duisburg-Essen, Lotharstraße 1, 47057 Duisburg, Germany} \label{UDE}
           \and
            {University Observatory, Faculty of Physics, Ludwig-Maximilians-Universit\"at M\"unchen, Scheinerstr.~1, 81679 Munich, Germany} \label{LMU}
            \and 
            {Exzellenzcluster ``Origins'', Boltzmannstr.~2, 85748 Garching, Germany} \label{garching}
            \and 
            {Institute for Computational Science, University of Z\"urich, Winterthurerstrasse 190, 8057 Z\"urich, Switzerland} \label{zuerich}
            \and 
            {ETH Z\"urich, Department of Physics, Wolfgang-Pauli-Strasse 27, 8093, Z\"urich, Switzerland} \label{ETH}}
          
   \date{Received 12/03/2025; accepted 25/07/2025}

 
  \abstract
  {}
  {
PDS\,70c is a source of H$\alpha$ emission and variable sub-millimetre signal. Knowledge of the emission mechanisms may enable observations of accretion rates and physical conditions in the circumplanetary environment.
}
{ We report  ALMA  observations of PDS\,70 at 145\,GHz (Band\,4), 343.5\,GHz (Band\,7), and 671\,GHz (Band\,9) and compare them with archival  data at 97.5\,GHz (Band\,3). The derived radio spectral energy distribution (SED) of PDS\,70c is coeval within two months, and is interpreted in terms of analytic models of dusty and viscous discs (i.e. circumplanetary discs, CPDs). In a novel approach, we include the free-free continuum from H\,{\sc i}, metals (e.g.\ K\,{\sc i}) and H$^{-}$.
}
   %
{ New detections in Bands~3 (tentative  at $2.6\,\sigma$), 4 ($5\,\sigma$), and 7 (re-detected at $9\,\sigma$) are consistent with optically thick thermal emission from PDS\,70c (spectral index $\alpha = 2\pm0.2$). However, a non-detection in Band\,9 breaks this trend, with a flux density falling below an optically thick extrapolation at $2.6\,\sigma$. A viscous dusty disc is inconsistent with the data, even with the inclusion of ionised jets.  Interestingly, the central temperatures in such  CPD models are high enough to ionise H\,{\sc i}, with huge emission measures and an optically thick spectrum that marginally accounts for the radio SED (within $3\, \sigma$ of  Band\,9). Since there is no room for steeper components (with $\alpha>2$), the dust-to-gas ratio is lower than $10^{-5}$.
By contrast, uniform-slab models suggest much lower emission measures to account for the Band\,9 drop, with ionisation fractions of $\sim 10^{-7}$ and an outer radius of $\sim 0.1\,{\rm au}$. Such conditions are recovered if the  CPD interacts with a  planetary magnetic field, leading to a radially variable viscosity, $\alpha(R)\lesssim1$, and central temperatures of $\sim$$10^{3}$\,K that regulate metal ionisation. However, the  $\Hm$ opacity still results in an optically thick SED, overshooting Band\,9. We find that the optically thin turnover at $\gtrsim$$600$ GHz is only recovered if a thin shocked layer is present at the CPD surface, as is suggested by  simulations.
A photospheric shock or accretion funnels are ruled out as radio emission sources because their small solid angles require  $T\sim$$10^{6}$~K, which are unrealistic temperatures in planetary shock accretion.
   }
   {The  SED of PDS\,70c collected here  is optically thick up to Band\,7 but probably (2.6\,$\sigma$) turns over towards Band\,9. An optically thick spectrum can be explained by atomic plasma radiation from a magnetised disc, where  the radio opacity stems from metal and $\Hm$ free-free. If so, PDS\,70c is depleted of sub-millimetre-emitting dust by a factor of at least 1000. However, the turnover can only be accounted for by H\,{\sc i} free-free from an accretion shock at the surface of a CPD.}

   \maketitle
%

   
\section{Introduction}


PDS\,70  hosts two protoplanets detected in direct imaging, and probably even three  \citep[][]{Keppler2018A&A...617A..44K,Haffert2019NatAs...3..749H, Wang2021AJ....161..148W, Christiaens2024A&A...685L...1C, Hammond2025MNRAS.539.1613H}. Atacama Large Millimeter Array (ALMA) observations of PDS\,70 resulted in the detection of a possible circumplanetary disc (CPD) around PDS\,70c \citep{Isella2019ApJ...879L..25I, Benisty2021ApJ...916L...2B}, in the form of  a point source in continuum emission at millimetre wavelengths in Band\,7 (351\,GHz), with a linear diameter of less than 1.2\,au, or about one third the size of the Hill radius of the accreting protoplanet. This CPD is likely to play a crucial role in regulating the accretion process of PDS\,70c. Its sub-millimetre flux density is variable by $\sim$40\% on timescales shorter than 2\,yr, and could be explained by the free-free continuum concomitant to the unresolved H$\alpha$ signal \citep[][]{CasassusCarcamo2022MNRAS.513.5790C} reported by \citep[][]{Haffert2019NatAs...3..749H}. For every proton-electron encounter leading to the recombination cascade, others will lead to free-free transitions.

The global hydrodynamic simulations of embedded thermal-mass bodies by \citet{Szulagyi2018MNRAS.473.3573S}, as well as the parametric viscous disc models by \citet{Zhu2018MNRAS.479.1850Z}, suggest that higher frequencies, such as those in ALMA Bands 9 and 10 ($\sim$671\,GHz and $\sim$850\,GHz), are more suitable for CPD observations in the dust continuum. Circumplanetary discs are predicted to be hotter than the surrounding circumstellar disc, and their (unresolved) emergent flux densities, $F_\nu$, in frequency, are thought to be mainly due to partially optically thick thermal dust with a spectral index of $\alpha \approx 3$ at ALMA frequencies, with $\alpha$ defined by $F_\nu\propto\nu^\alpha$.

However, if the radio signal from PDS\,70c is mainly due to free-free emission, we should expect $\alpha \leq 2$. A rough approximation of partially thick free-free emission can be obtained with a $1/r^2$ electron-density distribution, as in winds from early-type stars. The emergent free-free flux densities then follow a power-law spectrum, $F_\nu \propto \nu^\alpha$,  with $\alpha \sim 0.6$ \citep[][]{WrightBarlow1975MNRAS.170...41W}.

The sub-millimetre detection of PDS\,70c thus raises the questions of what the nature of the emission is, and why PDS\,70c is detected in the sub-millimetre continuum, while PDS\,70b is not. If the emission is due to thermal dust emission, a multi-wavelength analysis could shed light on the nature of the circumplanetary  dust population. In turn, the free-free continuum component will inform us directly about the planetary accretion process, without the large uncertainties related to the levels of extinction that affect H$\alpha$ (e.g., \citealp{Hashimoto2020AJ....159..222H}).

In this study, we present new ALMA Band\,4 (145\,GHz), Band\,7 (343\,GHz), and Band\,9 (671\,GHz) observations of PDS\,70, and compare them to archival data in Band\,3 data (97.5\,GHz) from \citet{Doi2024ApJ...974L..25D}, focussing on PDS\,70c. Section~\ref{sec:obs} describes data acquisition, synthesis imaging, and the extraction of the spectral energy distribution (SED) of PDS\,70c. Sec.~\ref{sec:analysis} analyses our results in terms of  parametric disc models. Sec.~\ref{sec:discussion} discusses which models can or cannot fit the data. Sec.~\ref{sec:conclusion} concludes.


\section{Observations}
\label{sec:obs}

\subsection{Data acquisition}
New ALMA observations of PDS\,70 in Bands\,4, 7, and 9 were obtained as part of ALMA project IDs 2022.1.00592.S and 2022.1.01477.S. Each band was acquired in an extended array configuration, labelled TM1, and a compact configuration, labelled TM2. An observation log is provided in Table~\ref{tab:observations}.  We also include in our analysis the archival datasets for Bands\,3, from ALMA projects 2022.1.00893.S \citep[][]{Doi2024ApJ...974L..25D} and 2022.1.01477.S \citep[][]{Liu2024ApJ...972..163L}.


\begin{table*}
  \caption{Observation log of the new Band\,4 and Band\,9 observations of PDS\,70.}             
  \label{tab:observations}      
  \centering                          
  \begin{tabular}{c c c c c c c c}        
    \hline\hline                 
    Band & UT Date  & ToS\tablefootmark{a}  & Elev.\tablefootmark{b}  &  PWV\tablefootmark{c}  & Phase\tablefootmark{d}   & Min BL\tablefootmark{e}   & Max BL \tablefootmark{f} \\    
    /dataset         &      & (s)   & (deg) &  (mm) &  rms (deg) &  (m)  &  (m) \\    
    \hline                        
    4/TM1 & 2023-07-09 & 5521 & 70.2 & 0.6 &  6.2 & 113 & 12752 \\
    4/TM1 & 2023-07-09 & 5485 & 66.3 & 0.6 &  5.6 & 113 & 12752 \\
    4/TM1 & 2023-07-10 & 5478 & 67.2 & 1.2 &  8.8 & 226 & 13814 \\
    4/TM1 & 2023-07-12 & 5511 & 66.3 & 0.6 &  8.8 & 226 & 15238 \\
    4/TM2 & 2023-04-30 & 3865 & 71.2 & 1.3 & 5.6 & 15 & 2517 \\  
    7/TM1 & 2023-05-23 & 10213 & 69 & 0.9 &  6.9 & 78 & 3638 \\
    7/TM1 & 2023-06-01 & 10128 & 57 & 0.5 &  6.3 & 27 & 3638 \\
    7/TM1 & 2023-06-02 & 9560 & 68 & 0.5 &  5.8 & 27 & 3638 \\
    7/TM1 & 2023-06-03 & 10109 & 62 & 0.7 &  10.3 & 27 & 3638 \\
    7/TM1 & 2023-06-04 & 9484 & 66 & 0.6 &  8 & 27 & 3638 \\
    7/TM1 & 2023-06-07 & 10113 & 66 & 0.6 &  8 & 77 & 3638 \\
    7/TM1 & 2023-06-08 & 10135 & 69 & 0.7 &  6.4& 27 & 3638 \\
    7/TM1 & 2023-11-27 & 10184 & 57 & 0.7 &  11.5& 31 & 3697 \\
    7/TM1 & 2023-11-28 & 10355 & 59 & 0.8 &  10.0& 31 & 3638 \\
    9/TM1 & 2023-06-11 & 7197 & 70.5 & 0.5 & 61 & 79 & 4614 \\
    9/TM2 & 2023-12-21 & 7171 & 68.4 & 0.5 & 52 & 15 & 1397 \\
    \hline                                   
  \end{tabular}
  \tablefoot{  
  \tablefoottext{a}{Time on source.}
  \tablefoottext{b}{Average elevation.}
  \tablefoottext{c}{Column of precipitable water vapor. }
  \tablefoottext{d}{Root-mean-square (rms) phase noise.}
  \tablefoottext{e}{Minimum baseline.}
  \tablefoottext{f}{Maximum baseline.}
  }
  
\end{table*}

\subsection{Imaging, self-calibration and alignment}
\label{sec:imaging}

\begin{figure*}
  \centering
  \resizebox{\hsize}{!}{\includegraphics{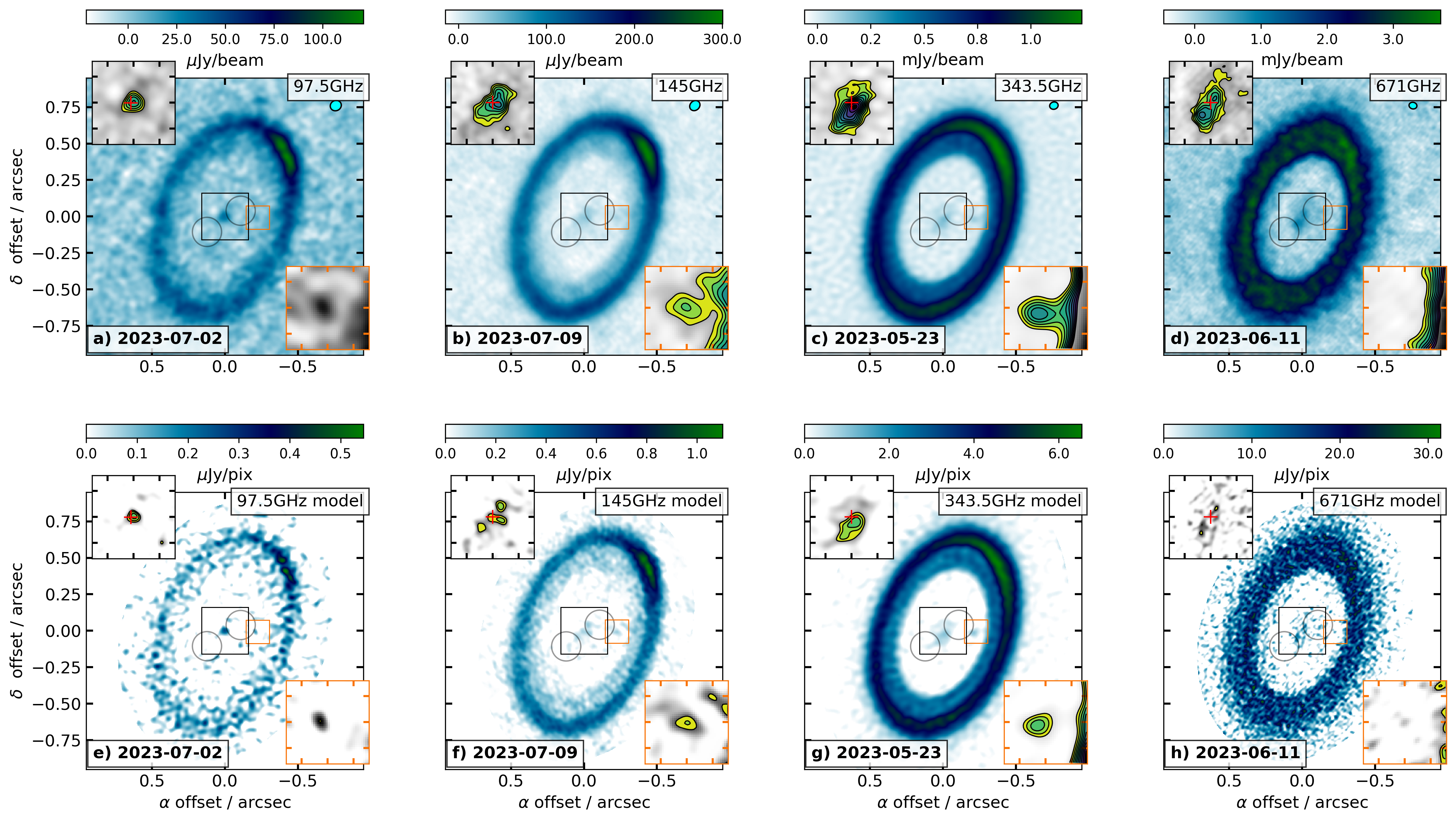}}
  \caption{Multi-frequency imaging of PDS\,70. The black insets at
    the top left of each image zoom on the central region (with
    tick-marks separated by $0\farcs1$), while the orange insets at
    the bottom right zoom on the expected position of PDS\,70c for
    circular Keplerian rotation around a $0.97\,M_{\odot}$ star (with
    tick-marks at $0\farcs05$). The red plus symbol in the central
    inset marks the nominal position of the star.  In all insets the
    linear grey scale stretches over the range of intensities in each
    region, and the contour levels start at $3\,\sigma$ and are
    incremented in units of $\sigma$. Intensities within each contour
    level are colour-coded differently.  The two circles are centred
    on the positions of PDS\,70b and PDS\,70d (no radio counterparts
    are detected).  Images in $a)$ to $d)$, along the top row,
    correspond to restorations of the corresponding non-parametric
    model images (that fit the visibility data), in $e)$ to $h)$,
    along the bottom row. A beam ellipse in $a)$ to $d)$ is shown in
    blue, on the top right. The pixel size in the model images is
    fixed at $4\times4\,$mas$^2$. For each band the Briggs parameter used for the restoration, the clean bean and the noise can be found in Table~\ref{tab:summary}. }
  \label{fig:summary}
\end{figure*}

The same imaging strategy as for the 2017 and 2019 Band\,7 datasets \citep[][]{CasassusCarcamo2022MNRAS.513.5790C,Casassus2023MNRAS.526.1545C} was extended to the other bands.  For each band and array configuration, automatic (user-independent) self-calibration was performed using the {\tt snow}  package. This package replaces the {\tt CASA} \texttt{tclean} imager \citep{Casa2022PASP..134k4501C} with \texttt{gpuvmem} \citep{Carcamo2018A&C....22...16C} in {\tt CASA} self-calibration (i.e. in iterative calls to the {\tt gaincal} and {\tt applycal} {\tt CASA} tasks). Image restoration was performed with a Briggs robustness parameter of \( r = 2 \).  The {\tt gaincal} solution interval was reduced by half in each iteration, starting with the scan length, and stopping when the peak signal-to-noise ratio (PS/N) decreases. The solution table with the highest PS/N was kept. After the phase calibration stage, a cycle of phase and amplitude self-calibration was performed to assess any PS/N improvements. During the antenna-gain calculations, using CASA task {\tt gaincal}, all spectral windows were combined. Alignment was performed using the {\tt VisAlign} package \citep[][]{CasassusCarcamo2022MNRAS.513.5790C,Casassus2023MNRAS.526.1545C}, selecting the datasets with the largest baselines as an astrometric reference (i.e. TM1), and a final round of self-calibration with {\tt snow} was applied to the concatenated dataset.  The resulting images are summarised in Fig.~\ref{fig:summary}.

For Band\,3, we concatenated the two extended (TM1) configurations from projects 2022.1.00893.S, 2022.1.01477.S. The PS/Ns of these data are limited by thermal noise, and self-calibration did not produce any improvements. The datasets were acquired with different phase centres, and were aligned with {\tt VisAlign} before concatenation. The reference was taken as 2022.1.00893.S.  The resulting Band\,3 continuum dataset yielded a PS/N of 13.6 with $r=2$, limited by thermal noise. As is shown in Fig.~\ref{fig:summary}a, a bright point source is detected at the phase centre, which is coincident with the position of PDS\,70  given by the GAIA ICRS co-ordinates \citep{Gaia2023A&A...674A...1G}. We therefore assume that this point source is due to accretion free-free emission on the central star, following \citet[][]{Doi2024ApJ...974L..25D}.

For Band\,4, each array configuration (TM1 and TM2) was self-calibrated separately prior to concatenation. The PS/N for TM2 improved from 224 to 243 after one round of phase self-calibration and one round of combined phase and amplitude self-calibration. In contrast,  TM1 is limited by thermal noise, with a PS/N of 49. An application of {\tt VisAlign}, with TM1 as a reference, showed that the pointing is very consistent in these datasets, with a  shift smaller than 1\,mas, and consistent intensities within 1\%. Such remarkable consistency probably reflects the favourable observing conditions. Joint self-calibration after concatenation did not yield improvements in the PS/N.

The new Band\,7 data is coarser than in \citet{Benisty2021ApJ...916L...2B}, with a restored beam in natural weights of $0\farcs102 \times 0\farcs082$ in TM1. We avoided concatenation with TM2 to preserve angular resolution. The PS/N of TM1 improves from 104 to 257 after two rounds of phase self-calibration, and reaches 308 after amplitude and phase self-calibration. However, we selected the phase-only self-calibration for our final dataset, as the amplitude gains result in a drop in flux density for PDS\,70 by 20\%, compared to both TM2 and the phase self-calibrated TM1.


In Band\,9, the PS/N for TM2 increases from 29.2 to 33.3 after two rounds of phase calibration, and to 111 after one round of combined amplitude and phase self-calibration. TM1 is affected by a low-spatial frequency modulation, probably caused by poor weather and resulting in noisy phases. We opted to discard baselines shorter than 0.4\,M$\lambda$. The uvrange-clipped dataset yielded a PS/N of 24, which did not improve with self-calibration. An application of {\tt VisAlign} to align TM2 to TM1 results in a shift of $\Delta \alpha $ = 25.6\,mas along the right ascension, and $\Delta \delta= 0.13$\,mas along the declination, with a correction in the intensity scale of 78\%. While seemingly large, the positional shift is within the pointing uncertainty of the TM2 dataset, which is about one tenth of the beam or 15\,mas (at $1\,\sigma$). Likewise, the absolute flux calibration accuracy in Band\,9 is around 10\%. Failing to apply the positional shift to TM2 results in images of the concatenated dataset with a spurious excursion of the ring into the cavity. We applied the correction factor of 1/78\% to TM1. Self-calibration of the concatenated dataset did not yield improvements.


While the star  PDS\,70 is readily picked up in  Band\,3, it bears no clear counterparts in the other bands. There may be a point-source associated with the star in Band\,4, where the central emission appears to be split into four blobs in the deconvolved  image of  Fig.~\ref{fig:summary}f, one of which coincides with the expected stellar position with $3\,\sigma$ flux. The other blobs are offset by $\sim 50$\,mas, which is too large to account for by a pointing error in Band\,4, given by $\sim 1/10$ of the Band\,4 TM1 beam, or $\sim 8$\,mas. We fitted elliptical Gaussians to extract the stellar flux densities. Comparing $24\pm5\,\mu$Jy in Band\,3 with $18\pm4\,\mu$Jy in Band\,4 corresponds to a spectral index of $\alpha = -0.7\pm0.8$, which is too noisy to distinguish partially thick free-free emission  \citep[$\alpha \sim 0.6$,][]{WrightBarlow1975MNRAS.170...41W},  variability, or a synchrotron component. We applied {\tt VisAlign} to check on the inter-band alignment, using  Band\,3 as reference given the detection of the star. Band\,4 is shifted by $\Delta \alpha = -4.9\,\pm1.1$\,mas, $\Delta \delta = -13\,\pm1.1$\,mas, which is  within the pointing errors in Band\,4.

There is no counterpart of the star in Bands\,7 or 9 with which to cross-check the inter-band alignment, and the ring appears dramatically different in radius and width compared to Band\,3, hampering the application of {\tt VisAlign}. We simply rely on the absolute pointing accuracy, of $\sim$10\,mas in Band\,7 and $\sim$5\,mas in Band\,9.


%

\begin{table*}
  \caption{Multi-frequency imaging properties of the restored images in Fig.~\ref{fig:summary}.}             
  \label{tab:summary}      
  \centering                          
  \begin{tabular}{c c c c c}        
    \hline\hline
    Bands       &   3    &   4   &   7   &   9  \\
    \hline
    $r$ \tablefootmark{a} & 1.0  & 1.0  & 0.0  & 2.0 \\
    $\Omega_b$ \tablefootmark{b}  & $0\farcs077\times 0\farcs070\,/\,-64$\,deg & $0\farcs075\times 0\farcs064 \,/\,-48$\,deg  & $0\farcs059\times 0\farcs050 \,/\,-75$\,deg  & $0\farcs054\times 0\farcs045 \,/\,78$\,deg  \\
    $\sigma$\tablefootmark{c} &  4.68  & 4.06  & 14  & 115  \\
    \hline                                   
  \end{tabular}
  \tablefoot{  \tablefoottext{a}{Briggs parameter used for the restoration.}
  \tablefoottext{b}{The beam major axis ({\tt bmaj}), minor axis ({\tt bmin}) and direction
    ({\tt bpa}) are in the format {\tt bmaj}$\times${\tt bmin}/{\tt bpa}.}
    \tablefoottext{c}{Noise, in $\mu$Jy\,beam$^{-1}$, in the restored images, as given by the root-mean-square dispersion in the imaging residuals.}}
\end{table*}

\subsection[Point source measurements of PDS70c]{Point source measurements of PDS\,70c}
\label{sec:pds70cdata}

\begin{table*}
  \caption{Point source measurements of PDS\,70c. }             
  \label{tab:PDS70c}      
  \centering                          
  \begin{tabular}{c c c c c c c}        
    \hline\hline                 
    Band & 3 & 4 & 7 &  7  &  7 & 9  \\    
    Date & 2023-07-02  & 2023-04-30  & 2017-12-02  &  2019-07-27  &  2023-05-23 & 2023-06-11    \\ 
    $\nu$/GHz & 97.5  & 145  & 350.1  & 350.1   &  343.5 &   671    \\ 
    \hline                        
    Flux / $\mu$Jy & $12\pm 4.7$ & $21.4\pm 4.1$  & $37\pm23.9$ & $118.5\pm16.6$ & $121\pm13$ & 143$\pm$115\tablefootmark{a} \\    
    $\delta$RA\tablefootmark{b} / mas & $-207.2 \pm 9$ & $-222 \pm 8$ &  ---   &  $-218\pm5$ & $-212.7 \pm 10.2$ &  --- \\
    $\delta$DEC\tablefootmark{c} / mas & $ -4.4 \pm 9$ & $ -5 \pm 8 $  & ---    & $27\pm5$ & $-12.1  \pm 10.2$ & --- \\
    \hline                                   
  \end{tabular}
  \tablefoot{%
  Uncertainties on the position (relative to PDS\,70) are taken as $1/10$ the clean beam major axis, in natural weights.
  \tablefoottext{a}{For Band\,9 we show a measurement of the intensity of the peak in a region centred on its expected position.}
  \tablefoottext{b}{Offset along right-ascension.}
  \tablefoottext{c}{Offset along declination.}
  }
\end{table*}

We searched for radio  counterparts to  PDS\,70\,b, c, and d by estimating their positions at the epochs of observations, under the assumption that their orbit is circular around a 0.97\,$M_\odot$ star, and in the plane of the disc, with position angle $\textrm{PA}=160\,$deg and inclination $i=130\,$deg \citep[e.g.][]{CasassusCarcamo2022MNRAS.513.5790C}. The two circles in each of the images in Fig.~\ref{fig:summary} are centred on the positions of b and d, using the co-ordinates provided by \citet[][from UT2020-02-10]{Wang2021AJ....161..148W} and \citet{Christiaens2024A&A...685L...1C}.

%
%
%
%
%
%
%
%

No counterparts were detected for PDS\,70b or d, at any frequency. However, we did find counterparts to PDS\,70c in Bands\,3 and 4. In Band\,3 a point-source signal at 2.6\,$\sigma$ in the restored map is found at $\Delta \alpha = 9\pm8$\,mas, $\Delta \delta = -1\pm8$\,mas, from the expected position of PDS\,70c, as was estimated with an elliptical Gaussian fit to the Band\,7 source (see Table~\ref{tab:PDS70c}). Given the pointing uncertainties in Band\,7 (of $\sim$5\,mas), the uncertainties in the orbit, and the faint Band\,3 counterpart that can be modulated by thermal noise, a shift of 9\,mas is well within the errors. We recorded the peak intensity in Jy\,beam$^{-1}$ as a measurement of the point-source flux, and set its uncertainty to the  noise in the residual image.

In Band\,4, a point source at 5.2\,$\sigma$ is found at $\Delta \alpha = 7 \pm 7$\,mas, $\Delta \delta = 2\pm7$\,mas from the expected position of PDS\,70c. The centroid of the  point source was estimated with an elliptical Gaussian fit in  the deconvolved image. Our choice of a Briggs parameter of $r=1$ in Sec.~\ref{sec:imaging} for image restoration is a compromise in terms of resolution to separate PDS\,70c from the outer ring and PS/N.


%

PDS\,70c is very conspicuous in the new Band\,7 data, especially in the deconvolved image of Fig.~\ref{fig:summary}g. The shift from its expected  position is  $\Delta \alpha = 13\pm12$\,mas, $\Delta \delta = -6\pm12$\,mas. However its flux density is difficult to measure because of confusion with the cavity wall. The peak on PDS\,70c, in the restored map of Fig.~\ref{fig:summary}c, is $111\pm14\,\mu$Jy, and a Gaussian fit to the deconvolved map of  Fig.~\ref{fig:summary}g yields $157\pm27\,\mu$Jy. We report the weighted average, $121\pm13\,\mu$Jy.

\begin{figure}
  \centering
\includegraphics[width=\columnwidth]{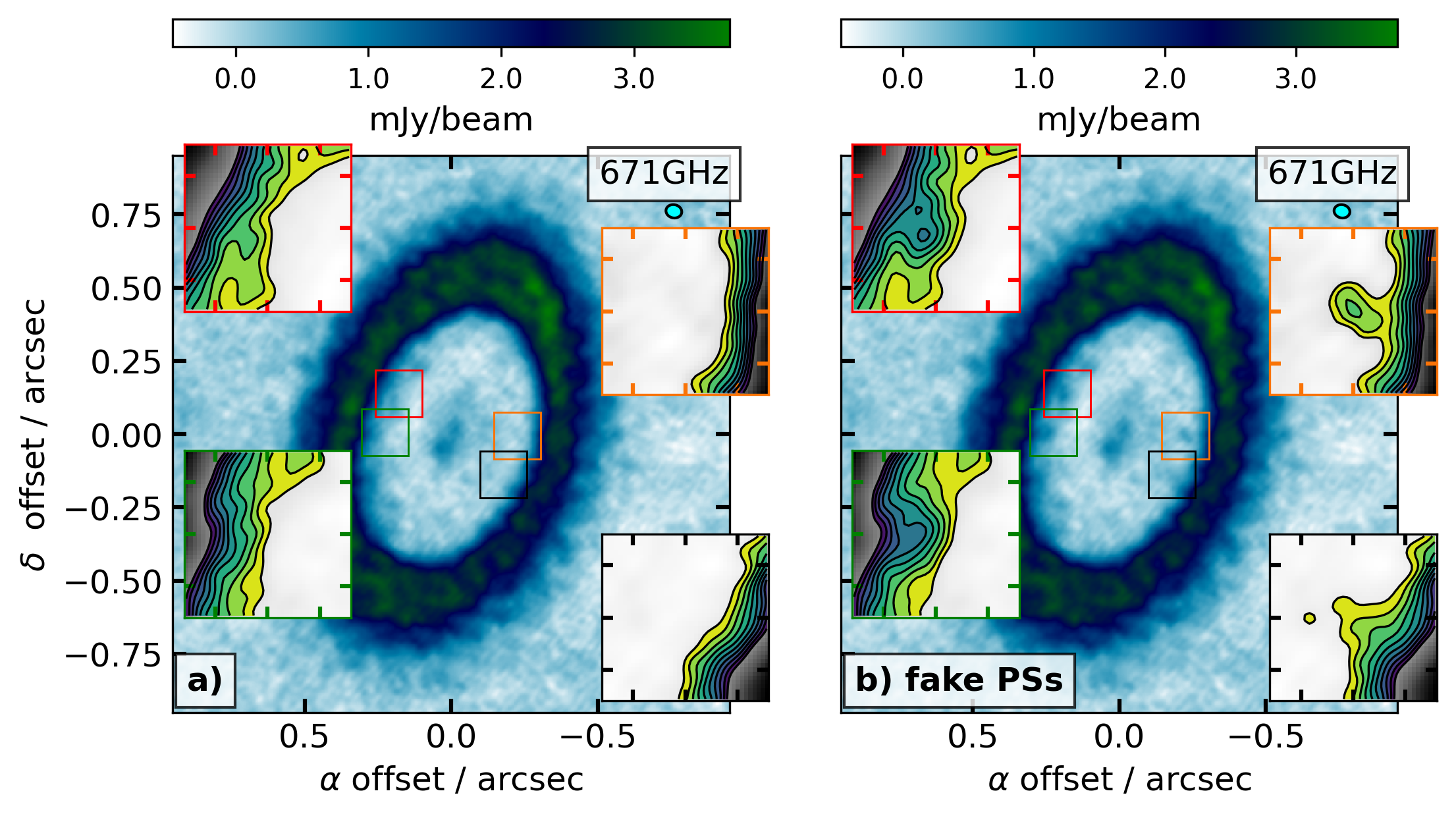}
  \caption{Point-source injections tests in Band\,9. a)~Same as Fig.~\ref{fig:summary}, with additional insets.  b)~Same as a) but with four point sources,  injected at the centre of each inset, and with a flux density corresponding to an optically thick extrapolation of the Band\,7 flux density for PDS\,70c.}
  \label{fig:PStests}
\end{figure}

Intriguingly, no counterparts of PDS\,70c were found in Band\,9. and the flux was obtained with the intensity of the peak in a region centred on its expected position, and with a radius given by 3 times the root-mean-square (rms) pointing accuracy (of $\sim 5$\,mas). We did not attempt to correct for the diffuse emission near the edge of the cavity, which would lower the flux density reported in Table~\ref{tab:PDS70c}. As the absence of PDS\,70c in Band\,9 is particularly surprising, we performed point-source injections  tests in the $uv$ plane, using optically thick spectra tied to Band\,7. As is shown in Fig.~\ref{fig:PStests}, such point sources should indeed be detectable, at least in the form of a protrusion of the cavity wall. 

The Band\,3 and Band\,9 flux density measurements of PDS\,70c presented here do not correspond to detections. However, their exclusion from the resulting SED would represent a severe loss of information. In turn, incorporating these non-detections as 2 or 3\,$\sigma$ upper limits injects a bias in attempts to fit this SED, by setting the expectation value for the flux densities to zero (or any other arbitrary value). The prior knowledge of the existence of PDS\,70c, and of its orbit, is incorporated by reporting the peak flux density, within a beam, at the expected location of PDS\,70c. This approach was referred to as `forced photometry'  by \citet{Samland2017A&A...603A..57S}, for the incorporation of   non-detections in the SED of 51\,Eri\,b.

%

With the flux measured in different bands, we use the spectral index $\alpha$ as a useful indicator of the emission mechanisms at radio frequencies,
\begin{equation}
  \alpha(\nu) = \frac{d \ln(F_{\nu})}{d \ln(\nu)}.
  \label{eq:spec_ind}
\end{equation}
We estimated $\alpha$ in PDS\,70c  using the measurements from Table~\ref{tab:PDS70c},
by comparing  frequency pairs, 
\begin{equation}
  \alpha_{\nu_1}^{\nu_2} = \frac{\ln (F_{\nu_2}/F_{\nu_1})}{\ln(\nu_2/\nu_1)},
  \label{eq:spec_ind_two_wav}
\end{equation}
and by fitting a power law to more than two frequency points,
\begin{equation}
  F_{\nu} = F_{\nu_\circ}   \left(\frac{\nu}{\nu_\circ}\right)^{\alpha},
  \label{eq:flux_spec_ind}
\end{equation}
with $\nu_\circ$ fixed to 145\,GHz.

A summary of different spectral index estimates is given in Table~\ref{tab:spec_ind}. Measurements involving Bands\,3, 4, 7, and 9 in 2023 are spread over two months (except for the November 2023 execution blocks in Band\,7 and the December 2023 TM2 execution block in Band\,9), and are the least biased by variability compared to the 2017 and 2019 data.

%
%

\begin{table*}
  \caption{Spectral indices for PDS\,70c.}             
  \label{tab:spec_ind}      
  \centering                          
  \begin{tabular}{c c c c c}        
    \hline\hline
    Bands       &   B7/B3     &  B7/B4  &   B7/B4/B3   &   B9/B7/B4/B3  \\
    $\alpha$    &  $1.84\pm0.32$ & $2.01\pm0.22$ &  $1.96\pm0.22$     &  $1.73\pm0.17$  \\
    \hline                                   
  \end{tabular}
\end{table*}

\subsection{Variability study}
\label{sec:var}  

\begin{figure*}
  \centering
\includegraphics[width=\linewidth]{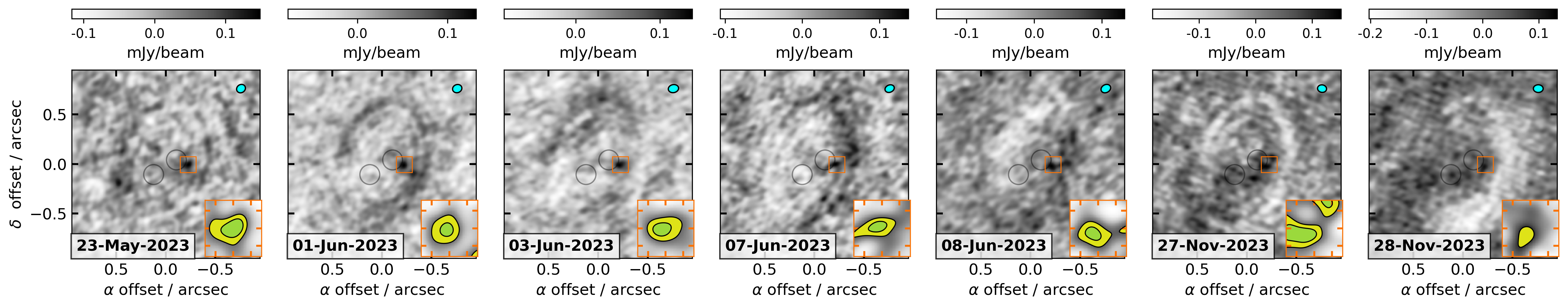}
\caption{Dirty maps in natural weights of the residual Band\,7 visibilities, after subtraction of the disc. The grey scale stretches over the full intensity scale, and for each of seven epochs. The inset zooms into PDS\,70c, with contours at 90 and 120\,$\mu$Jy\,beam$^{-1}$, fixed for ease of comparison across epochs. As in Fig.~\ref{fig:summary}, the circles are centred on PDS\,70b and d, and are drawn here to ease the search for any variable counterparts. }
  \label{fig:PDS70cresiduals_epochs}
\end{figure*}

The new Band\,7 data were acquired in nine execution blocks spanning seven epochs, and each of these execution blocks reached thermal noise levels of $\sim$20--30\,$\mu$Jy\,beam$^{-1}$,  in natural weights, which should be deep enough to detect PDS\,70c. However, the natural weight beams, of $\sim$0\farcs12, are too coarse to separate PDS\,70c from the bright ring. We therefore proceeded to subtract the disc emission from the visibility data. In this case we opted to work on the visibility data after amplitude and phase self-calibration, which ensures a consistent flux scale across all epochs. This was ascertained by aligning all epochs to 23 May 2023 with {\tt VisAlign}, resulting in positional shifts of less than 1\,mas, and amplitude shifts of less than 1\%. We prepared a disc image with the {\tt gpuvmem} model of the resulting visibilities (same as in Fig.~\ref{fig:summary}g), and subtracted PDS\,70c with an elliptical Gaussian fit. Simulated visibilities on this model image were then computed with the {\tt pyralysis} package \citep[][]{CasassusCarcamo2022MNRAS.513.5790C} and subtracted from the observations. The dirty maps of the residuals are shown in Fig.~\ref{fig:PDS70cresiduals_epochs}, in natural weights (and also computed with {\tt pyralysis}). The residual intensities are not perfectly thermal, as the model image fits the averaged dataset, not individual executions blocks. Nevertheless, PDS\,70c stands out as the brightest residual in all epochs.

The technique of $uv$-plane subtraction of the disc signal allowed us to build visibility data containing mostly the signal from PDS\,70c, with a consistent flux scale. We could thus proceed with point-source fits in the $uv$ plane, fixed at $\vec{x}_\circ$, the position of the elliptical Gaussian centroid used to subtract PDS\,70c, but with the best-fit amplitude (obtained by weighted least squares minimisation):

\begin{equation}
F_{\nu} = \frac{\sum_k \omega_k \Re\left[ V_k^\circ {\exp{\left(-2\pi i \,\vec{u}\cdot\vec{x_\circ} \right)}} \right]  }{ \mathcal{A}_\circ  \sum_k \omega_k},
\end{equation}
where $\Re[\cdot]$ denotes the real part, and with rms uncertainties
\begin{equation}
\sigma\left(F_{\nu}\right) = \frac{1}{\mathcal{A}_\circ \sqrt{\sum_k \omega_k}},
\end{equation}
where the sums extend over all observed visibilities $V_k^\circ$, at the frequency, $\nu$, with weights, $\omega_k$, and with a primary beam attenuation of $\mathcal{A}_\circ = \mathcal{A}(\vec{x}_\circ)$ at the position,  $\vec{x}_\circ$. We stress that while this technique is suitable for a variability study, it cannot be used to extract the flux density of the point source in the concatenated dataset, since the model image fits this point source.

\begin{figure}
  \centering
\includegraphics[width=0.8\columnwidth]{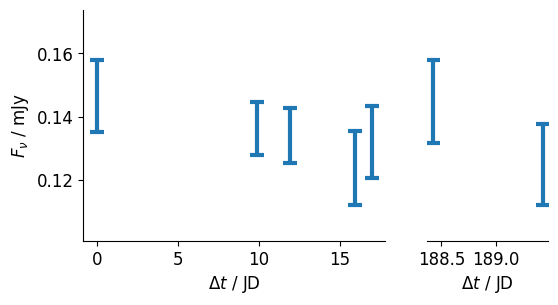}
  \caption{Flux density of PDS\,70c at 343.5\,GHz (Band\,7), as a function of Julian day. The scatter is consistent with the uncertainties.}
  \label{fig:Fdens_epochs}
\end{figure}

In practice we collapsed all channels in each spectral window, yielding four flux-density measurements. The weighted average flux density at each epoch is recorded in Fig.~\ref{fig:Fdens_epochs}.  The weighted average is $130\pm 4~\mu$Jy, with a  weighted scatter of 9.3\%. Therefore the flux density of PDS\,70c is constant across all seven epochs within $\sim$10\%. A non-detectable variability of the Band\,7 flux densities the PDS\,70c in 2023 is also consistent, given the uncertainties,  with the conclusion from \citet{Fasano2025A&A...699A.373F}, who compared measurements in 2019, 2021, and 2023. The Band\,4--Band\,7 spectral index derived here is consistent with the Band\,6--Band\,7 index reported by \citet{Fasano2025A&A...699A.373F}.



\section{Analysis}
\label{sec:analysis}

From the spectral indices in Table~\ref{tab:spec_ind}, we note that optically thick emission, that is $\alpha \gtrsim 2$, is consistent with the data between Bands\,3 and 7 but only marginally consistent when we add Band\,9. This is very difficult to reconcile with dust emission, where the emergent total (unresolved) flux densities mix optically thick and thin regions, resulting in spectral indices larger than 2.5 and closer to 3, depending on the choice of grain population  \citep[][]{Zhu2018MNRAS.479.1850Z,Shibaike2025ApJ...979...24S}. In fact, a blackbody fit  to Bands 3, 4, and 7 predicts $465\pm42\,\mu$Jy in Band\,9, and overshoots the observation by 2.6\,$\sigma$.

The non-detection of PDS\,70c in Band\,9 is intriguing, as CPD models predict a bright signal at such short wavelengths, and favour Band\,9 for CPD searches \citep[][]{Szulagyi2018MNRAS.473.3573S,Zhu2018MNRAS.479.1850Z}. It seems unlikely that this non-detection is caused by variability, given Sec.~\ref{sec:var}, and since PDS\,70c is readily detected in Bands\,4 and 7 at $\sim 5\,\sigma$ and $\sim 9\,\sigma$ respectively. Instead, the multi-frequency data suggest that free-free emission, rather than the thermal dust component,  may be the source of the PDS\,70c radio signal. In this section we analyse our results in terms of simple parametric models of the radio signal from CPDs, which we use to reproduce the measurements from Bands\,3, 4, 7, and 9.

\subsection[PDS70c SED models]{PDS\,70c SED models}

The lack of CPD detection in Band\,9 may be indicative of free-free radiation. To test this hypothesis, we extended  the parametric CPD model proposed by  \citet{Zhu2018MNRAS.479.1850Z} to include  the free-free emission expected from the CPD itself (here we consider the radio free-free contributions from all species). 
For completeness we reproduce here some of the formulae from \citet{Zhu2018MNRAS.479.1850Z}, before incurring some approximations and including our extensions. 

The model of \citet{Zhu2018MNRAS.479.1850Z} incorporates H\,{\sc i} free-free emission in the form of a jet. 
For this  component, we used the formula from \citet{Anglada2018A&ARv..26....3A} to estimate the emission of a circumstellar disc at 1\,cm:
\begin{equation}
  \left( \frac{F_{1\,{\rm cm},\,\rm jet} d^2}{\rm mJy~ kpc^2}\right)= 10^{-1.9} \left( \frac{L_{\rm acc}}{L_{\rm \odot}} \right)^{0.6},
  \label{eq:anglada_jet}
\end{equation}
with
\begin{equation}
  L_{\rm acc}=\frac{G M_{\rm p} \dot{M}_{\rm p}}{2 R_{\rm in}},
  \label{eq:bolometric}
\end{equation}
where $M_{\rm p}$ is the mass of the planet, $\dot{M}_{\rm p}$ the accretion rate onto it, and $R_{\rm in}$ the inner edge of the CPD.

The 1\,cm flux density of the jet may be extrapolated to millimetre wavelengths under the hypothesis of a power-law spectrum.  If the jet is partially optically thick, an upper limit is set on the spectral index with $\alpha_{\rm jet}=0.4$ \citep{Carrasco2016ApJ...821L..16C}. In the case of an optically thin jet, $\alpha_{\rm jet}=-0.1$. It should be borne in mind that these formulae were developed for stellar jets, rather than protoplanets, and that they are being extrapolated into the planetary mass regime.

The parametric CPD model from \citet{Zhu2018MNRAS.479.1850Z}  calculates the radio emission based on the disc accretion rate, $\dot{M}_{\rm p}$, and viscosity parameter, $\alpha$. Prior to applying this model to PDS\,70c, we made sure  to  reproduce the tables from \citet{Zhu2018MNRAS.479.1850Z} for the boundary layer case (this assumes that the planet accretion luminosity dominates over the central brightness of the planet itself), and in the Rayleigh-Jeans approximation for consistency.

We opted for a planetary mass of $M_{\rm p}=4 \, M_{\rm Jup}$, given the upper limit of $4.9 \, M_{\rm Jup}$ \citep[][based on a circumstellar disc viscosity parameter of
$\alpha=10^{-3}$]{Doi2024ApJ...974L..25D} and the restriction of $4 \, M_{\rm Jup}<M_{\rm p}< 12 \, M_{\rm Jup}$ \citep{Haffert2019NatAs...3..749H}. PDS\,70 is at a distance of $112.3 \, {\rm pc}$ \citep{Gaia2023A&A...674A...1G}, and we adopted a planetary orbital radius for PDS\,70c of $34 \, {\rm au}$  \citep{Haffert2019NatAs...3..749H}, implying an outer CPD radius of $R_{\rm out}=1/3\,\RHill \sim 1.23 \, {\rm au}$ \citep{Quillen1998ApJ...508..707Q}, corresponding to an angular size of $\Theta=0\farcs011$. The radius of the planet, $R_{\rm p}$, and the inner radius, $R_{\rm in}$, were both set to $1 \, R_{\rm Jup}$ following \cite{Zhu2018MNRAS.479.1850Z}.  This choice does not affect our results significantly since, as we shall see for the CPD models, most of the flux does not come from the inner region. 


The posterior distributions on the free parameters of our SED models were sampled with the  {\tt Nautilus} package \citep{Lange2023MNRAS.525.3181L}, which implements nested sampling aided by neuronal networks. We assumed flat priors in logarithm, with very wide boundaries.

We considered four models for the observed SED. Firstly, we considered only the dust component from the CPD. Secondly, we included an optically thin jet. Thirdly, we included an optically thick jet. Finally, we included free-free emission from the disc itself, using an $\alpha$-constant disc and a magnetic disc. We also guided our analysis using a uniform slab model, adding the contribution of free electrons from metals in the magnetic disc.

\subsection{Dust component of the CPD}
\label{sec:dust_comp}

\begin{figure}
  \centering
  \resizebox{\hsize}{!}{\includegraphics{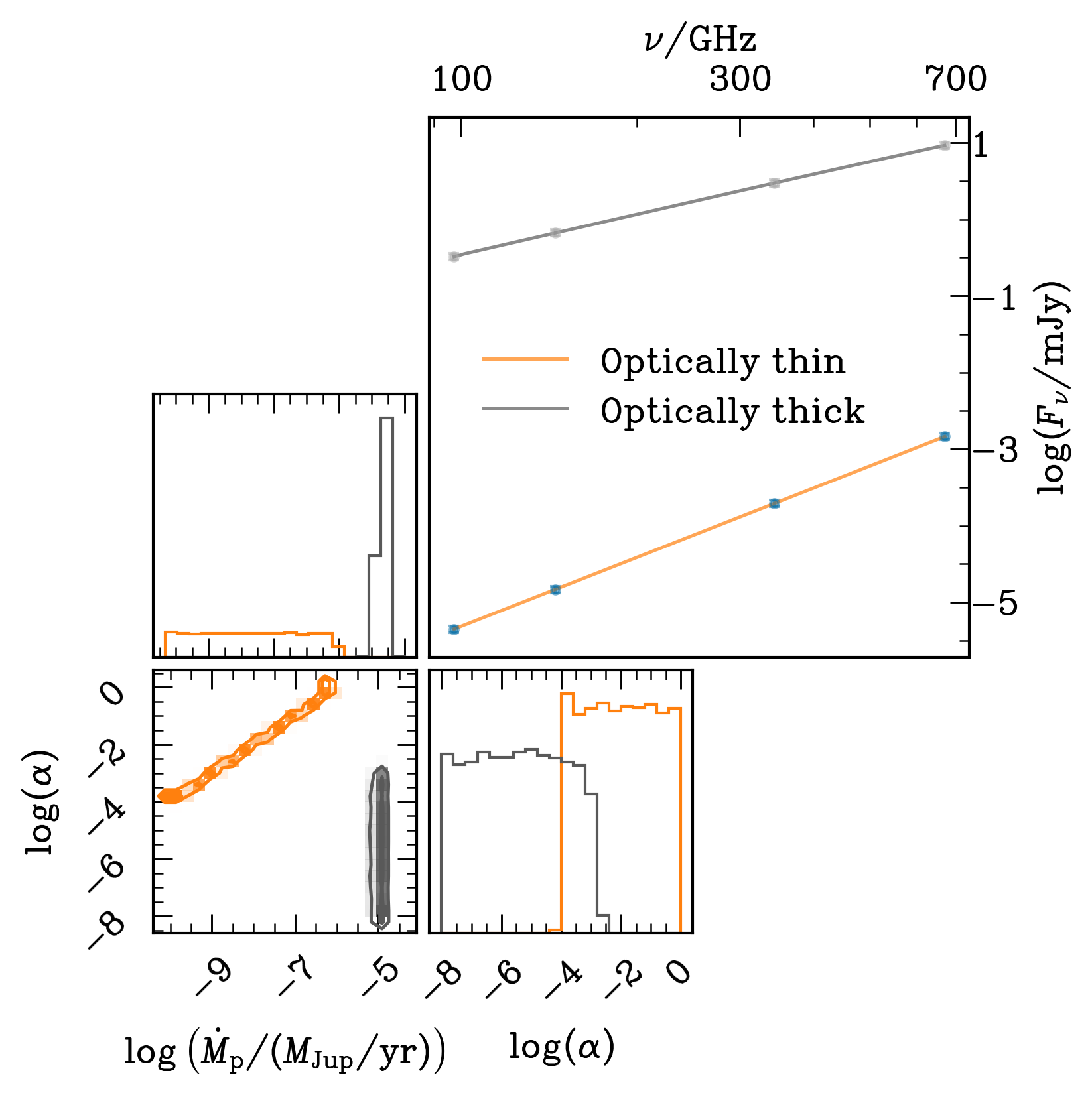}}
  \caption{Corner plot for synthetic data. The orange posteriors show the synthetic optically thin case, calculated with $\dot{M}_{\rm p}=10^{-9} \, M_{\rm Jup}\, {\rm yr}^{-1}$ and  $\alpha=10^{-3}$. The outcome illustrates the degeneracy between $\dot{M}_{\rm p}$ and $\alpha$ inherent to the parametric CPD model of \citet{Zhu2018MNRAS.479.1850Z}. The grey posteriors show the synthetic optically thick case, calculated with $\dot{M}_{\rm p}=10^{-5} \, M_{\rm Jup}\, {\rm yr}^{-1}$ (and the same $\alpha=10^{-3}$). Both synthetic SEDs were obtained using only the dust component of the CPD model following \citet{Zhu2018MNRAS.479.1850Z}, with errors set to 10\% of the flux.}
  \label{fig:corner}%
\end{figure}

In the absence of a jet, a strong degeneracy was observed between the parameters for smaller values of the accretion rate, $\dot{M}_{\rm p}$, and larger values of the viscosity parameters, $\alpha$. For example, using synthetic data obtained with the parametric model of \citet{Zhu2018MNRAS.479.1850Z}, for Bands\,3, 4, 7, and 9 with $\dot{M}_{\rm p}=10^{-9} \, M_{\rm Jup}\, {\rm yr}^{-1}$ and  $\alpha=10^{-3}$, we obtained the orange corner plot shown in Fig.\,\ref{fig:corner}, revealing that the important parameter is the ratio between these two parameters.
By contrast, in the case of a larger accretion rate, $\dot{M}_{\rm p}=10^{-5} \, M_{\rm Jup}\, {\rm yr}^{-1}$, and the same viscosity parameter, the flux depends solely on the accretion rate, as can be seen in the grey corner plot shown in Fig.~\ref{fig:corner} \citep[as noted by][]{Zhu2018MNRAS.479.1850Z}.

The degeneracy between $\dot{M}_{\rm p}$ and $\alpha$ can be understood by writing the emergent flux density, 
\begin{equation}
  F_{\lambda}=\frac{1}{d^2}\int_{R_{\rm in}}^{R_{out}} B_{\nu}(T_{\rm b}) 2 \pi R \,{\rm d}R,
  \label{eq:flux}   
\end{equation}
where $B_{\nu}$ is the Planck function (we use the full blackbody function rather than the  Rayleigh--Jeans approximation because the disc could reach low temperatures), $d$ is the distance to PDS\,70, and the brightness temperature is given by \citet{Zhu2018MNRAS.479.1850Z},
\begin{equation}
  T_{\rm b}=
  \begin{cases}
    \left( \frac{3}{8} \frac{\kappa_{\rm R}}{\kappa_{\rm mm}} T_{\rm eff}^4 + T_{\rm ext}^4 \right)^{1/4}  & {\rm if}~ \tau_{\rm mm}>0.5,
    \\
    \\
    2 T_{\rm c} \tau_{\rm mm} & {\rm if}~\tau_{\rm mm}<0.5.
  \end{cases}
  \label{eq:Tb}
\end{equation}
The Rosseland mean opacity was chosen to be constant, at the nominal value  $\kappa_{\rm R} = 10 \, {\rm cm}^{2}/ {\rm g}$ \citep[as in ][]{Zhu2018MNRAS.479.1850Z}. The dust opacity is  $\kappa_{\rm mm}=3.4 \zeta \times \frac{0.87 \rm mm}{\lambda}  \, {\rm cm}^{2}/ {\rm g}$  \citep[][with $\zeta$ the dust-to-gas ratio usually set as 0.01]{Andrews2012ApJ...744..162A}, and the optical depth $\tau_{\rm mm}=\kappa_{\rm mm} \Sigma / 2$. The effective temperature $T_{\rm eff}$ is the temperature that the surface of the viscous disc would radiate if it were a blackbody \citep[][]{Hartmann1998apsf.book.....H,Zhu2018MNRAS.479.1850Z},
\begin{equation}
  T_{\rm eff}^{4}=\frac{3 G M_{\rm p}\dot{M}_{\rm p}}{8 \pi \sigma R^3} \left[1-\left(\frac{R_{\rm in}}{R} \right)^{1/2} \right],
  \label{eq:Teff}
\end{equation}
and the external temperature is $T_{\rm ext}^4=T_{\rm floor}^4+T_{\rm irr}^4$, where  $T_{\rm floor}$ is the background temperature in the environment of the CPD and $T_{\rm irr}$ is the temperature-equivalent irradiation flux from the planetary photospheric luminosity. 

The background temperature in the circumstellar disc environment surrounding the CPD, $T_{\rm floor}$, is a crucial parameter as it sets the floor temperature in the CPD. We used $T_{\rm floor}=27 \, {\rm K}$, which was obtained by extrapolating the dust physical conditions in the outer ring obtained by multi-frequency SED fits (these dust diagnostics will be documented in a forthcoming article). The azimuthally averaged temperature profile of 21\,K at a de-projected radius of 0.5\,$\arcsec$ (56\,au) was extrapolated to the position of PDS\,70c at 0.3\,$\arcsec$ (i.e. 34\,au) under the assumption of a temperature profile of $R^{-1/2}$. In practice, the uniform-slab dust diagnostic  suggest a much higher temperature, as a power-law fit to the dust temperatures in the outer ring yields 49.9\,K at 34\,au. However, we adopted the value of $27 \, {\rm K}$ in a conservative approach, as a lower value widens the range of possible dust temperatures in the CPD. For comparison, the peak brightness temperature in Band\,9, of 14.7\,K at 67\,au, would translate to 21\,K at 34\,au with a $R^{-1/2}$ temperature profile. Additionally, the mid-plane temperature estimates by \citet{Law2024ApJ...964..190L} suggest  a temperature of at least 30\,K at the distance of PDS\,70c.

When the planetary accretion luminosity dominates over the central brightness of the planet we find that the irradiation temperature is 
\begin{equation}
  T_{\rm irr}=\left( \frac{L_{\rm acc}}{\sigma 40 \pi R^2} \right)^{1/4},
  \label{eq:Tirr}
\end{equation}
if the disc receives one tenth of the irradiation luminosity in the direction perpendicular to the disc \citep[i.e. as in][]{Zhu2018MNRAS.479.1850Z}.

The optically thick and thin regimes in $\tau_{\rm mm}$  can be distinguished from Eq.~\ref{eq:Tb}. In the optically thick regime, since $T_{\rm eff} \propto \dot{M}_{\rm p}^{1/4}$ and $T_{\rm ext} \propto \dot{M}_{\rm p}^{1/4}$ this implies that $T_{\rm b} \propto \dot{M}_{\rm p}^{1/4}$ (except for large radii where $T_{\rm ext} \sim T_{\rm floor}$). This results in a millimetre-flux density depending solely on the accretion rate, $F_\nu \propto \dot{M}_{\rm p}^{1/4}$.

In the optically thin regime, $T_{\rm b} \propto T_{\rm c} \Sigma$. The mass column density,  $\Sigma$, was calculated following \citet{Zhu2018MNRAS.479.1850Z}, as the minimum between the equations
\begin{equation}
  \Sigma= \frac{2^{7/5}}{3^{6/5}} \left( \frac{\sigma G M_{\rm p} \dot{M}_{\rm p}^3}{\alpha^4 \pi^3 \kappa_{\rm R} R^3}\right)^{1/5}  \left(\frac{\mu}{\mathfrak{R}} \right) \left[ 1- \left( \frac{R_{\rm in}}{R}  \right)^{1/2} \right]^{3/5}  \propto \left( \frac{\dot{M}_{\rm p}^3}{\alpha^4} \right)^{1/5} ,
  \label{eq:sigma_visc_heat}
\end{equation}
and
\begin{equation}
  \Sigma = \frac{\dot{M}_{\rm p} \mu \Omega}{3 \pi \alpha T_{\rm ext}} \propto \left( \frac{\dot{M}_{\rm p}^3}{\alpha^4} \right)^{1/4},
  \label{eq:sigma_ext}
\end{equation}
where $\mathfrak{R}$ is the ideal gas constant and $\mu$ is the mean molecular weight set as $\mu=1$. The mid-plane temperature is given by \citep[][]{Zhu2018MNRAS.479.1850Z}:
\begin{equation}
  T_{\rm c} = \left( \frac{9 G M_{\rm p} \dot{M}_{\rm p} \Sigma \kappa_{\rm R}}{128 \pi \sigma R^3}  \left[ 1- \left( \frac{R_{\rm in}}{R}  \right)^{1/2} \right] + T_{\rm ext}^{4} \right) ^{1/4} .
  \label{eq:Tc}
\end{equation}
If we neglect the contribution of $T_{\rm ext}$, which is for a purely viscously heated disc, then we keep $\Sigma$ from Eq.~\ref{eq:sigma_visc_heat},  and Eq.~\ref{eq:Tc} yields
\begin{equation}
  T_{\rm c} \Sigma = \sqrt{\frac{G M_{\rm p}}{\pi^2 R^3}} \frac{\dot{M}_{\rm p}}{\alpha} \left( \frac{\mu}{\mathfrak{R}} \right)^{5/4} \left[ 1 - \left( \frac{R_{\rm in}}{R}  \right)^{1/2} \right].
  \label{eq:Tc_sigma}
\end{equation}
Finally from Eq.~\ref{eq:flux} if the disc becomes optically thin at a sufficient small radii, which is the case with $\tau_{\rm mm}<0.5$ of Eq.~\ref{eq:Tb}, $T_{\rm b}$ can be calculated using Eq.~\ref{eq:Tc_sigma} since most of the contribution will come from larger radii (optically thin case).

The following simplifications can be made if the accretion rate is small enough. First, neglect the planetary accretion luminosity relative to the viscous heating mechanism with $T_{\rm ext}\sim 0$. Second, approximate that the CPD is mostly optically thin by lowering the surface density, $\Sigma \propto \left(\dot{M}_{\rm p}^3/\alpha^4 \right)^{1/5} $. A CPD heated by viscous dissipation allows for a simplified model of the SED, with optically thin millimetre emission that depends on a single parameter,
\begin{equation}
\gamma=\frac{\dot{M}_{\rm p}/(M_{\rm Jup}/ {\rm yr})}{\alpha}.
\label{eq:gamma}
\end{equation}


\subsection[HI free-free component of the disk]{H\,{\sc i} free-free component of the CPD} \label{sec:free-free_comp}

Another component of the SED from PDS\,70c is free-free emission from the disc itself. This scenario is supported by mid-plane temperatures exceeding \(10^4\)~K in the steady-state CPD models of  \citet{Zhu2018MNRAS.479.1850Z}.  Also, the hydrogen recombination leading to the H$\alpha$ signal from PDS\,70c \citep[][]{Haffert2019NatAs...3..749H} implies that the radio signal is, at some level, partly due to    H\,{\sc i} free-free continuum emission.

Here we consider the H\,{\sc i} free-free luminosity of the parametric disc models from \citet{Zhu2018MNRAS.479.1850Z}. The surface density $\Sigma(R)$ (Eqs.~\ref{eq:sigma_visc_heat} and \ref{eq:sigma_ext}), gives information on the gas density. Only the regions hot enough to ionise hydrogen ($T \gtrsim 10000 \,{\rm K}$) will contribute to the flux, and we neglect free electrons contributed by other elements.


Assuming that the gas follows a hydro-static and Gaussian vertical profile, 
\begin{equation}
  \rho (R,z) = \frac{\Sigma(R)}{\sqrt{2 \pi} H(R)} \exp \left( -\frac{z^2}{2 H(R)^2} \right),
  \label{eq:density}
\end{equation}
with the scale height $H(R)=c_{\rm s}/\Omega(R)$. We solved the radiative transfer equation with vertical stratification of the temperature \citep[following ][]{Hubeny1990ApJ...351..632H},
\begin{equation}
  T(z)^{4} = \frac{3}{8} T_{\rm eff}^{4} \tau_{\rm R}+T_{\rm ext}^{4},
\end{equation}
where $\tau_{\rm R}$ is the Rosseland mean optical depth at z. We can calculate the number density of electrons using the Saha equation 
\begin{equation}
  \frac{n({\rm H\,\textsc{ii}})}{n({\rm H\,\textsc{i}})}=\frac{2 {\rm Q_\textsc{ii}}}{n_{\rm e}{\rm Q_{\textsc{i}}}} \left( \frac{2 \pi m_{\rm e} \kB T}{h^2} \right)^{3/2} \exp \left(-\frac{\chi}{\kB T}\right),
  \label{eq:saha}
\end{equation}
where  $m_{\rm e}$ is the electron mass,  $\kB$ is the Boltzmann constant, $\chi$ is the collisional ionisation potential of hydrogen (13.6\,eV), $h$ is the Planck constant, and the partition function was calculated as
\begin{equation}
  {\rm Q_{\textsc{i}}}=\sum_{n=1}^{3} 2 n^2 \exp{\left(-{\frac{\chi(1-1/n^2)}{\kB T}}\right)},
  \label{eq:partition}
\end{equation}
for neutral hydrogen, and ${\rm Q_\textsc{ii}}=1$ for ionized hydrogen. The ionisation fraction of an initially neutral gas is given by $f = {n_{\rm e}}/{n_{\rm H}}$, where $n_{\rm H}=\rho(R,z)/m_{\rm H}$ (we assume the gas consists of hydrogen only). We can then solve for $n_{\rm e}$,
\begin{multline}
  n_{\rm e}(R,z)= \frac{\rm Q_{\textsc{ii}}}{\rm Q_{\textsc{i}}}\left( \frac{2 \pi m_{\rm e} \kB T}{h^2} \right)^{3/2} \exp \left( \frac{-\chi}{\kB T} \right) \\ \left(-1  + \sqrt{ 1 + 2 n_{\rm H}  \frac{\rm Q_{\textsc{i}}}{\rm Q_{\textsc{ii}}}  \left( \frac{2 \pi m_{\rm e} \kB T}{h^2} \right)^{-3/2} \exp \left(\frac{\chi}{\kB T}\right)}  \right).
  \label{eq:n_e}
\end{multline}

With the number density of electrons we calculated the emission measure (EM),
\begin{equation}
  {\rm EM}(R,z)=\int_{z}^{\infty} n_{\rm e}(R,z)  n_{\rm i}(R,z) dz,
\end{equation}
where $n_{\rm i}$ is number density of positive ions (for this initially neutral gas $n_{\rm i}=n_{\rm e}$). We also calculated the optical depth \citep[][including stimulated emission]{WrightBarlow1975MNRAS.170...41W,Allen1973asqu.book.....A}, 
\begin{equation}
  \tau_{\nu}(R, z)=\int_{z}^{\infty} K_{\nu}(T) n_{\rm e}(R,z)  n_{\rm i}(R,z) dz,
  \label{eq:tau}
\end{equation}
with
\begin{multline}
  K_{\nu}(T)=\frac{4\pi}{3\sqrt{3}}\frac{Z^2e^6}{hcm_{\rm e}^2\varv}\frac{g_{\nu}}{\nu^3} \left(1-\exp \left(-\frac{h \nu}{\kB T} \right) \right)\\=\frac{3.692 \times 10^{8}}{\rm cm^{-5}\, K^{-1/2}\, Hz^{-3}}\left(1-\exp \left(-\frac{h \nu}{\kB T}\right)\right) Z^2 g_{\nu}T^{-1/2} \nu^{-3},
\end{multline}
where the electron velocity $\varv$ was replaced by the mean velocity $\bar{\varv}=(\pi k T/2m_{\rm e})^{1/2}$, $Z$ is the ionic charge (in this case $Z=1$), $e$ is the electron charge and $c$ is the velocity of light in vacuum. The gaunt factor, $g_{\nu}$, was approximated using the equation from \citet{Oster1961RvMP...33..525O}
\begin{multline} 
  g_{\nu}(T)=\ln\left({\left(\frac{2 \kB T}{\gamma m_{\rm e}}\right)^{3/2}\left(\frac{2 m_{\rm e}}{\gamma Z e^2 \omega}\right)}\right) \\ =\ln\left(4.955 \times 10^7 \left(\frac{\nu}{\rm Hz}\right)^{-1}\right)+1.5 \ln\left(\frac{T}{\rm K}\right), \label{eq:gaunt}
\end{multline}
with $\gamma=e^{\gamma^{*}}\approx 1.781$ the exponential of Euler's constant $\gamma^{*}$ and $\omega=2 \pi \nu$ the angular frequency. When $g_{\nu}(T) < 1$, we set $g_{\nu}(T) = 1$, which is the limit at higher frequencies.
\footnote{There is a typo in the formula for the Gaunt factor in  \citet[their Eq.~25]{2000beckertA&A...356.1149B}. Here  we use  the constants from \citet{Oster1961RvMP...33..525O}.}

Finally, we calculated the free-free emission from the CPD by integrating over the disk
\begin{equation}
  F_{\nu}=\int_{R_{\rm in}}^{R_{\rm out}} \int_{-\infty}^{\infty} e^{-(\tau_{\nu}-\tau_{\nu}')} B_{\nu}(T(z') )K_{\nu}(T(z'))  n_{\rm e}^{2}(R,z') dz' \, \frac{ 2 \pi R}{d^{2}} dR,
  \label{eq:F_ff}
\end{equation}
where we have already replaced $n_{\rm i}=n_{\rm e}$, $B_{\nu}(T)$ is the Planck function, $\tau_{\nu} = \tau_{\nu} (\rm R, z \to - \infty)$, is the total optical depth from Eq.\,\ref{eq:tau}, and $\tau_{\nu}' = \tau_{\nu} (\rm R, z')$, is the  optical depth calculated using Eq.\,\ref{eq:tau}, integrating from $z'$ to $\infty$. The distance to PDS\,70 is $d$.

\subsection{Uniform slab model}
\label{sec:uniform_slab}

In parallel to the calculation of the free-free component from the analytic viscous disc model, we also implemented a simpler treatment in order to inform  ourselves about the EMs and disc sizes that would match the observed SED. We used a pill-box model consisting of a uniform slab with a constant temperature and EM, and a cross-section of $A= \pi R_{\rm max}^2$. The optical depth given by Eq.~\ref{eq:tau} reduces to
\begin{equation}
  \tau_{\nu}^{\rm pill}= K_{\nu}(T) \times {\rm EM},
  \label{eq:tau_pill}
\end{equation}
then from Eq.~\ref{eq:F_ff}, since $T$ and EM are constants, we solved both integrals analytically using $R_{\rm in}=0$ and $R_{\rm out}=R_{\rm max}$ and obtained
\begin{equation}
  F_{\nu}^{\rm pill}=\pi B_{\nu}(T) \left(1-e^{ -\tau_{\nu}^{\rm pill}} \right) \left( \frac{R_{\rm max}}{d} \right)^{2}.
  \label{eq:F_pill}
\end{equation}

\subsection{Magnetic disc}
\label{sec:mhd_disk}

The previous models did not take into account the free electrons coming from metals, since the viscous disc models in consideration reach hydrogen-ionising temperatures. However, the very high effective viscosity in a strongly magnetised disc would result in lower peak temperatures, as in the parametric disc models from  \citet{Hasegawa2021ApJ...923...27H}. Such `magnetic disc' models assume that the planet is undergoing magnetospheric accretion, which means that the inner region of the CPD has enough free electrons to couple the gas to the magnetic field of the planet.

We followed \citet{Hasegawa2021ApJ...923...27H}, with a magnetic field ($B_{\rm p}$) described as a dipole
\begin{equation}
  B_{\rm p}= B_{\rm ps} \left(\frac{R_{\rm p}}{R} \right)^3,
  \label{eq:Bps}
\end{equation}
where $B_{\rm ps}$ is the  field strength at the surface of the planet.
We used an alpha-viscosity ($\alpha_{\rm m}$) inspired by the results of the ideal magnetohydrodynamics (MHD) simulations of \citet{Salvesen2016MNRAS.457..857S},
\begin{equation}
  \alpha_{\rm m} = 11\, \beta^{-0.53},
  \label{eq:alpha_MHD}
\end{equation}
where $\beta=\rho_{\circ}c_{\rm s}^2/(B_{\rm p}^2/8 \pi)$  and $\rho_{\circ}=\Sigma \Omega/(\sqrt{2 \pi}c_{\rm s})$ is the mid-plane gas volume density.

Since the dusty disc model from Sec.~\ref{sec:dust_comp} already has the disc surface density, $\Sigma(R)$, and the temperature profile, $T_{\rm c}(R)$, for a given $\alpha$ and $\dot{M}_{\rm p}$, we can solve Eq. ~\ref{eq:alpha_MHD} by imposing $\alpha=\alpha_{\rm m}$ for a fixed $B_{\rm ps}$ and $\dot{M}_{\rm p}$. Thus, the magnetic disc model has a variable viscous parameter $\alpha(R)$, with the rest of the physical conditions calculated analogously to Sec~\ref{sec:dust_comp}. However, these viscosity estimates neglect other sources of ionisation  (i.e. interstellar UV and cosmic rays), and $\alpha$ values can be unrealistically low at large radii. We therefore imposed a lower limit, $\alpha(R) > 10^{-5}$.


With the temperature and density profiles as a function of $\dot{M}_{\rm p}$ and $B_{\rm ps}$, the next step was to consider the free electrons coming from the metals. For this we used the abundances and collisional ionisation potentials of the most relevant elements in the solar photosphere, shown in Table~\ref{tab:metals} \citep[][]{Asplund2009ARA&A..47..481A,Lide2008chcp.book.....L,Sarah2014MNRAS.440...89K}.

\begin{table}
  \caption{Atomic number (Z), abundance and collisional ionisation potential ($\chi_{\rm i}$) of the most relevant elements.}             
  \label{tab:metals}      
  \centering                          
  \begin{tabular}{c c c c}        
    \hline\hline
    Element       &  Z  &  Abundance  & $\chi_{\rm i}$ (eV) \\
    \hline
    H & 1 & $9.21 \times 10^{-1}$ & 13.60 \\
    He & 2 & $7.84 \times 10^{-2}$ & 24.59 \\
    Na & 11 & $1.60 \times 10^{-6}$ & 5.14 \\
    Mg & 12 & $3.67 \times 10^{-5}$ & 7.65 \\
    K & 19 & $9.87 \times 10^{-8}$ & 4.34 \\
    \hline                                   
  \end{tabular}
\end{table}

We used these abundances and solved the Saha equation for each element, imposing the neutrality condition
\begin{equation}
  n_{\rm e} = \sum_{i}n_{\rm i},
  \label{eq:neutrality}
\end{equation}
where the sum runs over the five elements in consideration, and $n_{\rm i}$ is the number density of the i-ionised element. We assumed that the metals are not depleted into dust.  However, solving this equation takes too long for an optimisation. We decided to use a simpler approach that takes advantage of the differences in abundance of each element. We assume that at a given temperature a single element contributes most of the free electrons. We solved the Saha equation separately, setting $n_{\rm e}=n_{\rm i}$ for each element, and used the abundances to determine which element provided most of the electrons at a given temperature \citep[a similar approach can be found in][]{Mohanty2018ApJ...861..144M}. This allows for an approximation of the real number density of electrons without requiring an excessively long computing time.

Following  \citet{Hasegawa2021ApJ...923...27H}, we take into account that the CPD is undergoing magnetospheric accretion. This means that the radius at which the planetary magnetic field dominates over the ram pressure (called the truncation radius $R_{\rm T}$) is larger than the radius at which the disc has enough free electrons to couple the gas to the magnetic field. This condition is satisfied for $T \sim 1000 \, \rm K$, which is the temperature threshold for
K, and we used the condition $R_{\rm T}>R(T_{\rm c}=1000 \, \rm K)$ to ensure that the model is consistent \citep{Hasegawa2021ApJ...923...27H}. Finally, we calculated the optical depth due to metal free-free using the electron number density in Eq.~\ref{eq:tau}.


\subsection{Negative hydrogen ion contribution}
\label{sec:H_minus}
The optical/IR opacity of the sun and cooler stars is dominated by the negative hydrogen ion H$^{-}$ \citep{Wildt1939ApJ....90..611W}, and it is worth exploring how it contributes to the CPD radio opacity. Because of its simple and highly polarised structure, the hydrogen atom can hold a second electron with an ionization potential of 0.7542~eV \citep{Frolov2015EPJD...69..132F}.


We included Wildt's photo-detachment, which is the bound-free mechanism,
\begin{equation}
  h \nu + {\rm H}^{-} \rightarrow {\rm H} + {\rm e}^{-},
  \label{eq:photo_detachment}
\end{equation}
where any photon with $\lambda<16440\,$\AA{} had sufficient energy to ionise the ${\rm H}^{-}$. We also included Pannekoek's free-free mechanism \citep{Pannekoek1931MNRAS..91..519P},
\begin{equation}
  h \nu + {\rm e}^{-} + {\rm H} \rightarrow {\rm H} + {\rm e}^{-},
  \label{eq:H_minus_ff}
\end{equation}
which contributes to the opacity, especially at longer wavelengths.

The presence of ionised metals is key to the abundance of ${\rm H}^{-}$, since we need extra electrons to encounter neutral hydrogen, and ${\rm H}^{-}$ also draws free electrons from the medium. This means we have to solve Saha's equation again adding the ${\rm H}^{-}$. We have its ionisation energy (0.7542~eV), partition function $\rm Q_{{\rm H}^{-}} \sim 1$, $\rm Q_{{\rm H}} \sim 2$ and a new neutrality condition
\begin{equation}
  n_{{\rm e}}+n_{{\rm H}^{-}} = \sum_i n_{\rm i},
  \label{eq:new_neutrality}
\end{equation}
which follows from Eq.~\ref{eq:neutrality}. The difference is that some of the free electrons from the ionised metals are now bound to the ${\rm H}^{-}$.

Here again we simplified the solution of the Saha equation by taking advantage of the different abundances of each element.  In most cases, the abundance of ${\rm H}^{-}$ was negligible compared with the ions, and we solved Saha's equation as shown in Sec~\ref{sec:mhd_disk}. We calculated the number density of ${\rm H}^{-}$  with the number density of electrons ($n_{\rm e}$) obtained with metals only. In the case in which the number density of ${\rm H}^{-}$ becomes significant (10\%\ of the ions or more),  the free electrons are less abundant in the medium. The number density of free electrons decreases drastically and we set $n_{\rm i}=n_{{\rm H}^{-}}=n_{{\rm e}}=0$.

Provided with the number density of electrons, we calculated the bound-free and free-free opacities of ${\rm H}^{-}$ following \citet[their Chap.~8]{Gray2022oasp.book.....G}. The final total optical depth is
\begin{equation}
  \tau_{\nu}^{\rm tot}=\tau_{\nu}^{\rm metals}+\tau_{\nu}^{{{\rm H}^{-}}_{\rm ff}}+\tau_{\nu}^{{{\rm H}^{-}}_{\rm bf}},
  \label{eq:total_tau}
\end{equation}
with $\tau_{\nu}^{\rm metals}$ the optical depth calculated as Sec.~\ref{sec:mhd_disk}, and $\tau_{\nu}^{{\rm H^{-}}_{\rm bf}}$ and $\tau_{\nu}^{{\rm H^{-}}_{\rm ff}}$ the bound-free and free-free optical depth of ${\rm H}^{-}$, respectively. The total flux was then calculated as
\begin{equation}
  F_{\nu}=\int_{R_{\rm in}}^{R_{\rm out}} \int_{0}^{\tau_{\nu}} e^{-(\tau_{\nu}-\tau_{\nu}')}B_{\nu}(\tau_{\nu}') d \tau_{\nu}' \, \frac{ 2 \pi R}{d^{2}} dR,
  \label{eq:F_tot}
\end{equation}
where $\tau_{\nu}$ is the $\tau_{\nu}^{\rm tot}$ calculated using Eq.~\ref{eq:total_tau}.

\subsection{Fitting the data}
\label{sec:fit_data}



\begin{figure}
  \centering
  \resizebox{0.8\hsize}{!}{\includegraphics{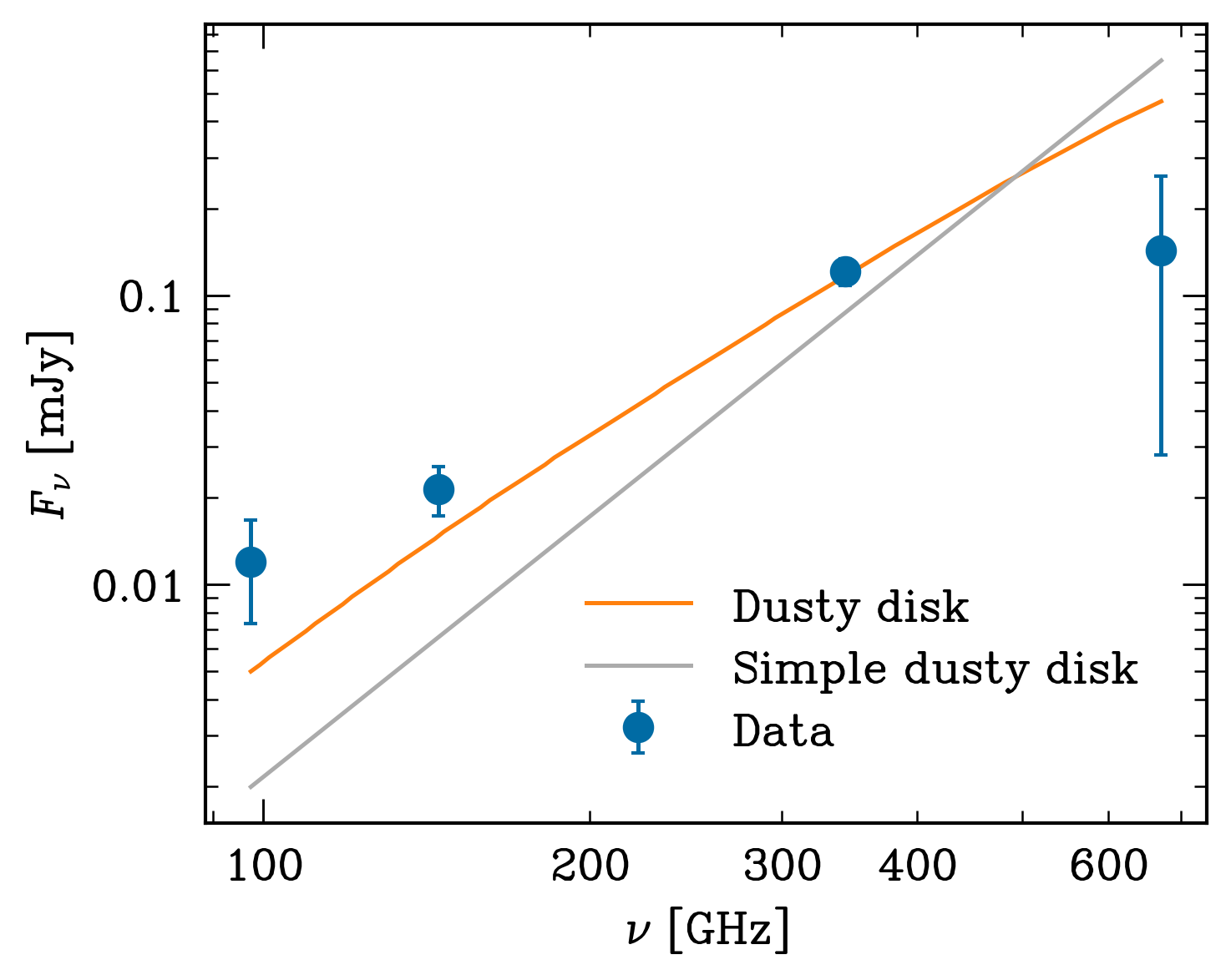}}
  \caption{Best fits for the data of Bands\,3, 4, 7, and 9 using the dusty disc \citep{Zhu2018MNRAS.479.1850Z} model and the simple dusty disc model (described in Sec.\,\ref{sec:dust_comp}) are shown as the solid orange and light grey lines, respectively. For the dusty disc, we find that the best fit parameters of the model are $\log(\dot{M}_{\rm p}/(M_{\rm Jup}/\rm yr))=-9.04^{+0.46}_{-2.75}$ and $\log(\alpha)=-5.21^{+0.53}_{-2.82}$ at $1\,\sigma$. The resulting spectral index is 2.35 and has a minimum reduced $\chi^{2}$ of 6.50. For the simpler dusty disc model, we find that the best fit parameter is $\log(\gamma)=-2.67_{-0.05}^{+0.04}$. The resulting spectral index is 3 and has a minimum reduced $\chi^{2}$ of 14.61.  Here the reduced $\chi^{2}$ is the $\chi^{2}$ divided by the number of points (data) minus the number of degrees of freedom in the model.}
  \label{fig:bestfitdust}%
\end{figure}

We first fitted the nearly coeval data in Bands\,3, 4, 7, and 9 using the dusty CPD model from Sec.\,\ref{sec:dust_comp}, varying both parameters ($\alpha$ and $\dot{M}_{\rm p}$). We observed that the accretion rate and viscosity parameters were degenerate, and therefore applied the simpler model that only depends on $\gamma$ (Eq.\,\ref{eq:gamma}). The best fit of both models is shown in Fig.~\ref{fig:bestfitdust}, which poorly fits the data.

\begin{figure}
  \centering
  \resizebox{0.8\hsize}{!}{\includegraphics{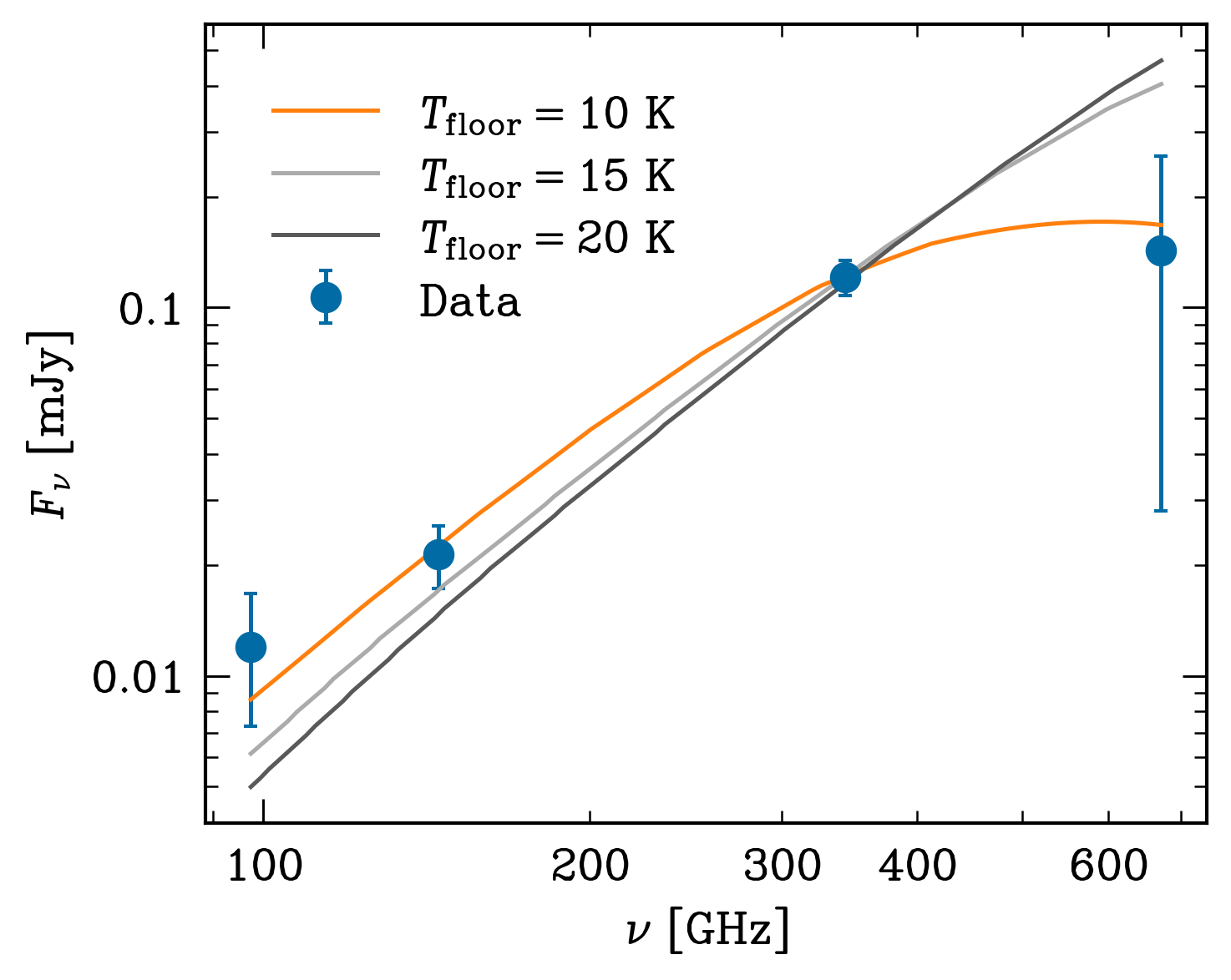}}
  \caption{Best fit for the data of Bands\,3, 4, 7, and 9 for the dusty disc \citep{Zhu2018MNRAS.479.1850Z} using different $T_{\rm floor}$. In order, the solid orange, light grey, and grey lines correspond to $T_{\rm floor}$ of 10, 15, and 20\,K.}
  \label{fig:bestfitzhu_T}%
\end{figure}

For the dusty disc model shown in Fig.~\ref{fig:bestfitdust}, we observed that the bulk of the flux originates from the outer radius, where the mid-plane temperature is $T_{\rm c}=T_{\rm ext} \approx T_{\rm floor}=27\,{\rm K}$. This indicates that the SED is highly dependent on the floor temperature, and our approximation using the single parameter, $\gamma$, was invalid ($T_{\rm ext}$ cannot be neglected). As is shown in Fig.~\ref{fig:bestfitzhu_T}, if the floor temperature were lower (i.e. $T < 15\,{\rm K}$), we could fit the data.

The addition of an optically thin and thick jet to the dusty disc model was performed. We fixed $\alpha=10^{-2}$ and also varied the dust/gas ratio, $\zeta$, using uniform priors with $\log(\gamma)$ and $\log(\zeta)$ between -15 and 15. However, in both cases the contribution of the jet after the fit was negligible and it just returned the dusty disc case without a jet.

\begin{figure}
  \centering
  \resizebox{0.8\hsize}{!}{\includegraphics{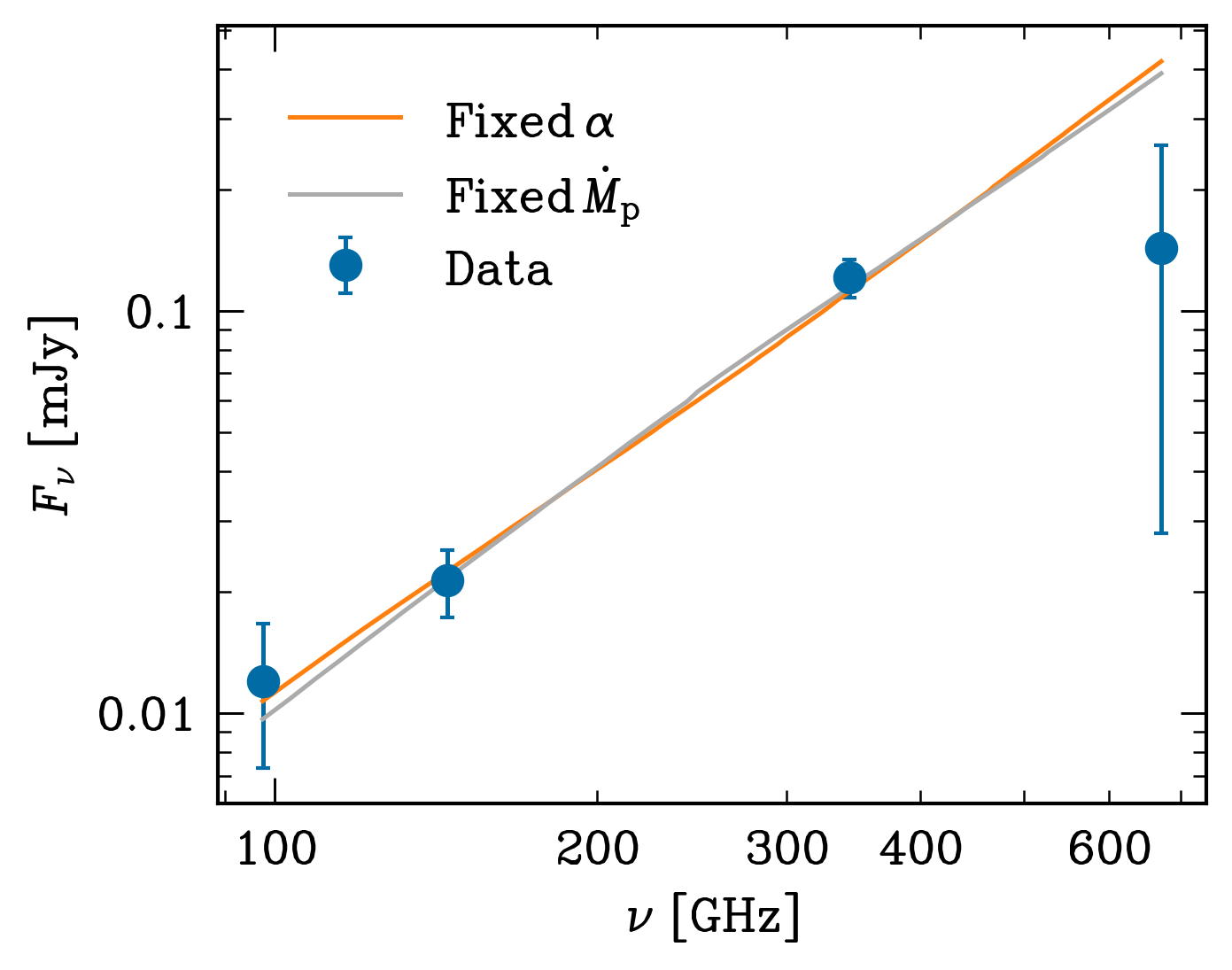}}
  \caption{Best fits to the data from Bands\,3, 4, 7, and 9 using two free-free emission models (sum the dusty disc component from Sec.\,\ref{sec:dust_comp} and the free-free component from Sec.\,\ref{sec:free-free_comp}). The first has a fixed $\alpha=10^{-2}$, shown as the solid orange line and has a minimum reduced $\chi^{2}$ of 3.18. We find that the best fit parameters of this model are $\log(\dot{M}_{\rm p}/(M_{\rm Jup}/\rm yr))=-3.48^{+0.06}_{-0.09}$ at $1\,\sigma$, with a strict upper limit on the dust-to-gas ratio of $\log(\zeta)=-4.80$. The second has a fixed $\dot{M}_{\rm p}=10^{-6} \, M_{\rm Jup} \,{\rm yr}^{-1}$ as the solid light grey line and has a minimum reduced $\chi^{2}$ of 2.54. We find that the best fit parameters for this model are $\log(\alpha)=-6.86^{+0.08}_{-0.07}$ at $1\,\sigma$, with a strict upper limit on the dust-to-gas ratio of $\log(\zeta)=-7.05$.}
  \label{fig:ff_zhu}%
\end{figure}

We then turned to the new model that sums the free-free component (from Sec.\,\ref{sec:free-free_comp}) and the thermal dust emission from the disc. Since there is no certainty that the approximation made in Sec.~\ref{sec:dust_comp} is valid (with the single parameter, $\gamma$), we used the complete model proposed by  \citet[][]{Zhu2018MNRAS.479.1850Z}. We considered two different models: fixed $\alpha$, for which we kept a fixed alpha value of $\alpha=10^{-2}$ and allowed the accretion rate to vary, and fixed $\dot{M}_{\rm p}$, for which we kept the accretion rate fixed at  $\dot{M}_{\rm p} = 10^{-6} \, M_{\rm Jup}\,{\rm yr}^{-1}$ and allowed the alpha value to vary. Fig.~\ref{fig:ff_zhu} shows the resulting best fits for each case. The models were able to produce spectral indices of  $(\alpha_{97.5}^{145},\alpha_{145}^{343.5}, \alpha_{343. 5}^{671}) = (1.87 , 1.86, 1.97)$ and $(\alpha_{97.5}^{145},\alpha_{145}^{343.5}, \alpha_{343.5}^{671})=(2.02, 1.93, 1. 83)$, respectively, which are more consistent with Table~\ref{tab:spec_ind} for Bands\,3 to 7. However, these models only fit the data points for Bands\,3, 4, and 7 and do not account for the non-detection in Band\,9 with a flux, $F_{671}$, of $417 \, \mu {\rm Jy}$ and $389 \, \mu {\rm Jy}$ in each case.

In the model with fixed $\alpha$ we derive a well-constrained mass accretion rate, and vice versa. However, the results also indicate that free-free emission is the dominant component of the CPD signal. The inferred upper limits on the dust-to-gas ratio are $10^{-4.80}$ and $10^{-7.05}$ for the models with fixed $\alpha$ and  fixed $\dot{M}_{\rm p}$, respectively. This suggests a strongly dust-depleted environment, which could point to reduced dust transport into the CPD, with larger grains being trapped in a pressure bump outside the planet's orbit \citep[e.g.][]{Zhu2012ApJ...755....6Z}. \citet{Weber2018ApJ...854..153W} show that in the presence of viscosity, even with strong gap-opening planetary torques, there is a flow of gas and micron-sized grains through the planet-induced gap, while grains larger than a certain size are filtered out at the gap edge. In contrast, a study more focussed on dust transport to the CPD by \citet{Szulagyi2022ApJ...924....1S} reports a contradictory result: the planet's gravitational perturbation elevates millimetre-sized particles from the mid-plane and induces a vertical accretion flow from the circumstellar disc towards the CPD. Since \citet{Weber2018ApJ...854..153W} considered a two-dimensional case over long evolutionary timescales -- leaving the planet sufficient time to clear its orbit -- whereas \citet{Szulagyi2022ApJ...924....1S} conducted three-dimensional simulations evolved for only 200 orbits, it remains an open question how dust filtration and transport operate in more evolved, three-dimensional cases.


The model with a fixed $\alpha=10^{-2}$ requires very high accretion rates, $\dot{M}_{\rm p} \sim 10^{-3.5} \, M_{\rm Jup}\, {\rm yr}^{-1}$. However, by reducing the viscosity parameter, for example to $\alpha=10^{-3}$, the flux ends up being overestimated by more than a factor of two. This means that reducing the viscosity parameter would require a lower accretion rate to fit the data. Using the same example, for $\alpha=10^{-3}$ the accretion rate that fits the data is $\dot{M}_{\rm p} \sim 10^{-4} \, M_{\rm Jup}\, {\rm yr}^{-1}$.

\begin{figure}
  \centering
  \resizebox{0.8\hsize}{!}{\includegraphics{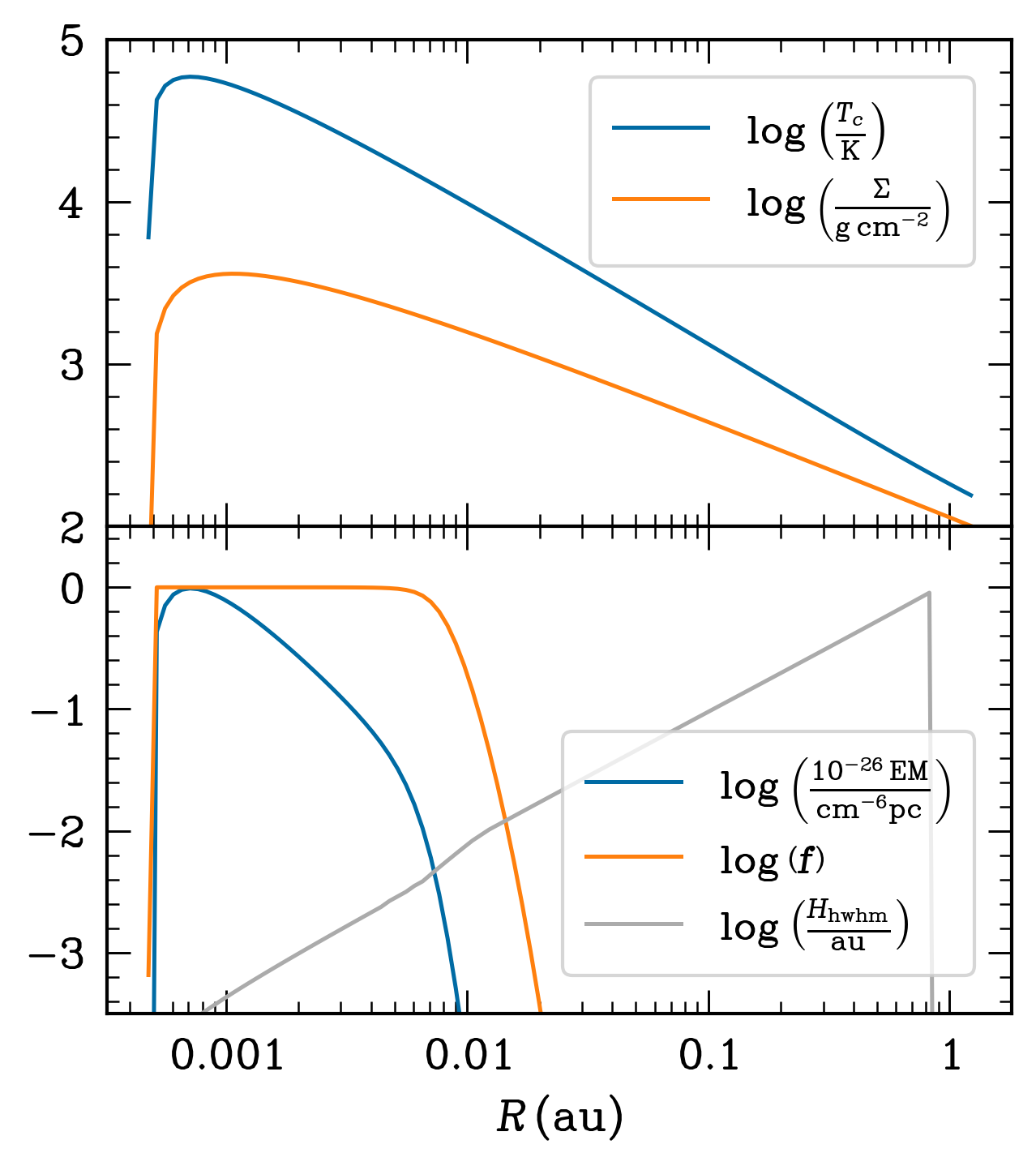}}
  \caption{Physical conditions including free-free in the best-fit CPD model with $\alpha$ fixed. $T_{\rm c}$ is the mid-plane temperature calculated as Eq.~\ref{eq:Tc}, $\Sigma(R)$ is the surface density, EM is the emission measure, $f$ is the ionisation fraction in the mid-plane, and $H_{\rm hwhm}$ is the height at which the density of free electrons is half the density of free electrons in the mid-plane (half-width at half-maximum).
  }
  \label{fig:profilesfixedalpha}%
\end{figure}

\begin{figure}
  \centering
  \resizebox{0.8\hsize}{!}{\includegraphics{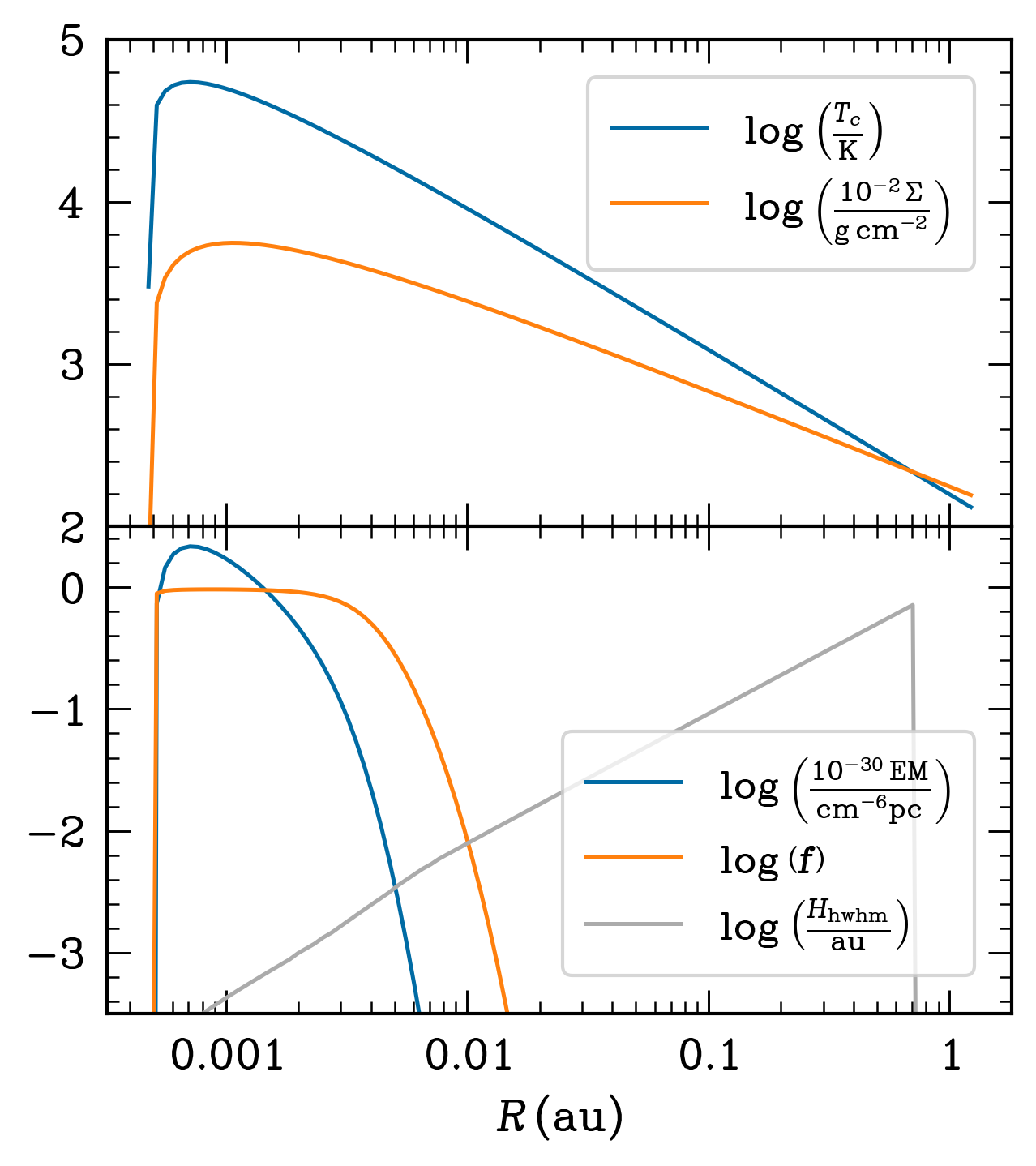}}
  \caption{Physical conditions in the best-fit CPD model with $\dot{M}_{\rm p}$ fixed. Annotations follow from Fig.~\ref{fig:profilesfixedalpha}.
  }
  \label{fig:profilesfixedMpdot}%
\end{figure}

The properties of the disc are listed in Figs.~\ref{fig:profilesfixedalpha} and \ref{fig:profilesfixedMpdot} for the case with fixed $\alpha$ and fixed $\dot{M}_{\rm p}$ respectively. At small radii ($R \lesssim 0.02 \rm au$) the fraction of ionisation is $f\sim 1$, and the resulting EMs are extremely large  ($\rm EM \sim 10^{26-31} \, {\rm cm}^{-6} \, {\rm pc}$). The profile for the surface density ($\Sigma$) is in the viscous-disc regime (small $T_{\rm ext}$), and in the case with fixed $\dot{M}_{\rm p}$  and small $\alpha$ (i.e. $\alpha \sim 10^{-6}$), $\Sigma$ is two orders of magnitude higher than for large values of $\alpha$ (i.e. $\alpha \sim 0.01$). Another outstanding feature of the physical conditions in these analytical viscously accreting CPDs is the occurrence of extremely hot inner discs, for any $\alpha < 0.01$. Such hydrogen-ionising temperatures are counter-intuitive for a CPD.

\begin{figure}
  \centering
  \resizebox{\hsize}{!}{\includegraphics{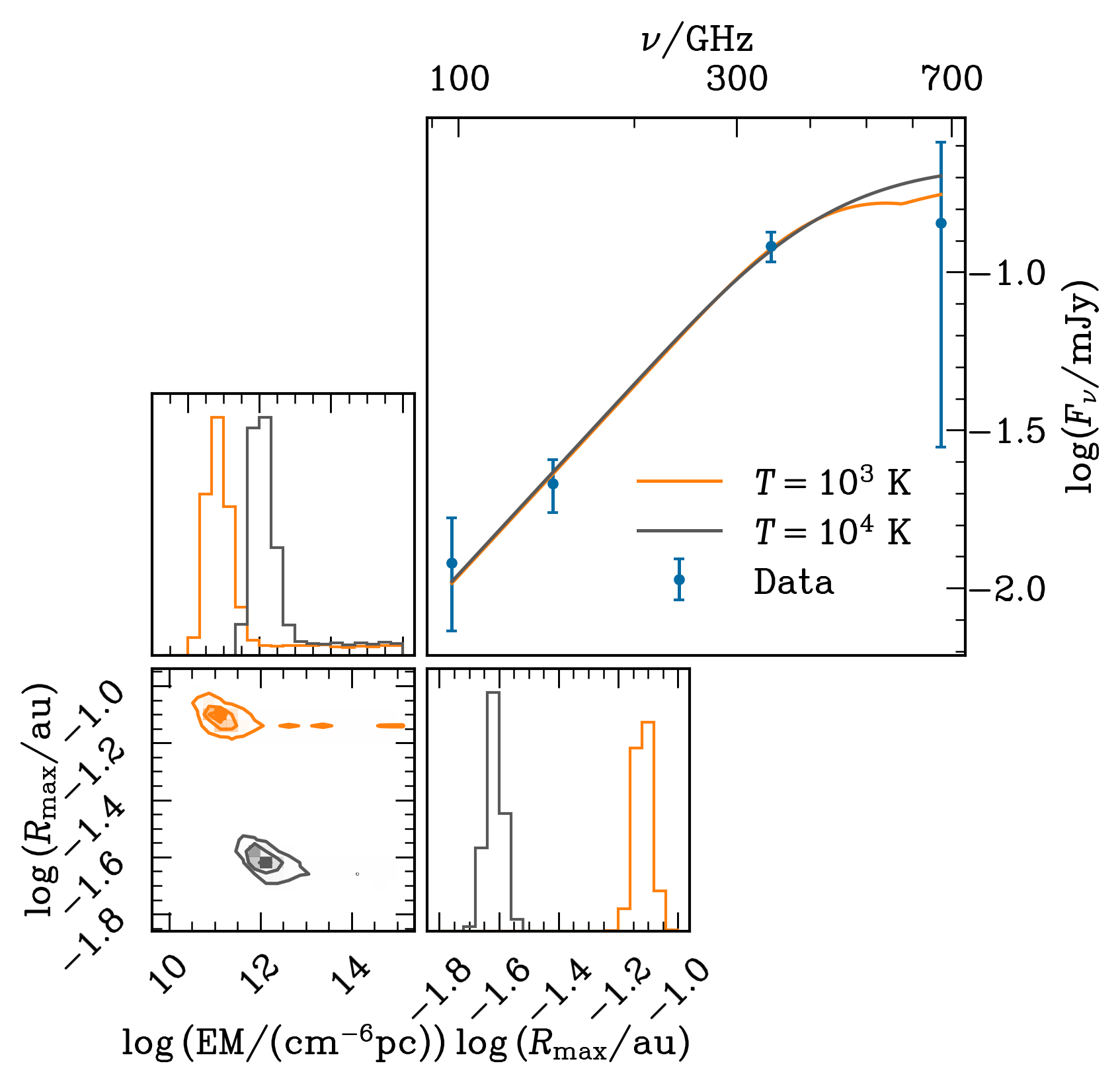}}
  \caption{Best fit to the data from Bands\,3, 4, 7, and 9 using the uniform slab model described in Sec.\,\ref{sec:uniform_slab}. On the upper right panel, the solid orange line corresponds to the case $T=10^{3}$\,K, has a minimum reduced $\chi^{2}$ of 0.18, and a p value of 83\%. We find that the best fit parameters of this model are $\log \left({\rm EM}/({\rm cm^{-6} \,pc}) \right)=11.08^{+0.53}_{-0.19}$ and $\log \left({R_{\rm max}}/{\rm au} \right)=-1.11^{+0.03}_{-0.06}$. The solid light grey line corresponds to the case $T=10^{4}$\,K, has a minimum reduced $\chi^{2}$ of 0.33, and a p value of 72\%. We find that the best fit parameters of this model are $\log \left({\rm EM}/({\rm cm^{-6} \,pc}) \right)=12.00^{+0.42}_{-0.18}$ and $\log \left({R_{\rm max}}/{\rm au}\right)=-1.61^{+0.02}_{-0.04}$. The bottom left panel shows the corner plots obtained for each model. The orange and grey posteriors also correspond to the cases of $T=10^3$\,K and $T=10^4$\,K, respectively. }
  \label{fig:pill}
 \end{figure}

For the uniform slab model, we fixed a temperature of $10^3\, \rm K$ and $10^4 \,\rm K$ and chose uniform priors for $\log\left(R_{\rm max}/ \rm au\right)$ and  $\log\left({\rm EM}/{(\rm cm^{-6}\, pc)}\right)$ (with very wide boundaries). The resulting best fits are shown in Fig.~\ref{fig:pill}. Now we could produce spectral indices of $(\alpha_{97.5}^{145},\alpha_{145}^{343.5}, \alpha_{343.5}^{671}) = (2.00, 1.91, 0.59)$ and  $(\alpha_{97.5}^{145},\alpha_{145}^{343.5}, \alpha_{343.5}^{671})=(2.00, 1.88, 0.81)$, respectively, which are more in line with Table~\ref{tab:spec_ind} even for Band\,9.

We notice that the values for the EM are smaller compared with the parametric model of \citet{Zhu2018MNRAS.479.1850Z}, having the best fit ${\rm EM}=1.19 \times 10^{11} \, {\rm cm}^{-6}\, {\rm pc}$ and ${\rm EM}=1.04 \times 10^{12} \, {\rm cm}^{-6}\, {\rm pc}$ for each case. On the other hand, the radius of the pill seems to match the order of magnitude of the radius where we start having a smaller ionisation fraction in Figs.~\ref{fig:profilesfixedalpha} and \ref{fig:profilesfixedMpdot}, with a best fit size of $R_{\rm max}=0.078 \, {\rm au}$ and $R_{\rm max}=0.025 \, {\rm au}$ for the pill in each case.



\begin{figure}
  \centering
  \resizebox{\hsize}{!}{\includegraphics{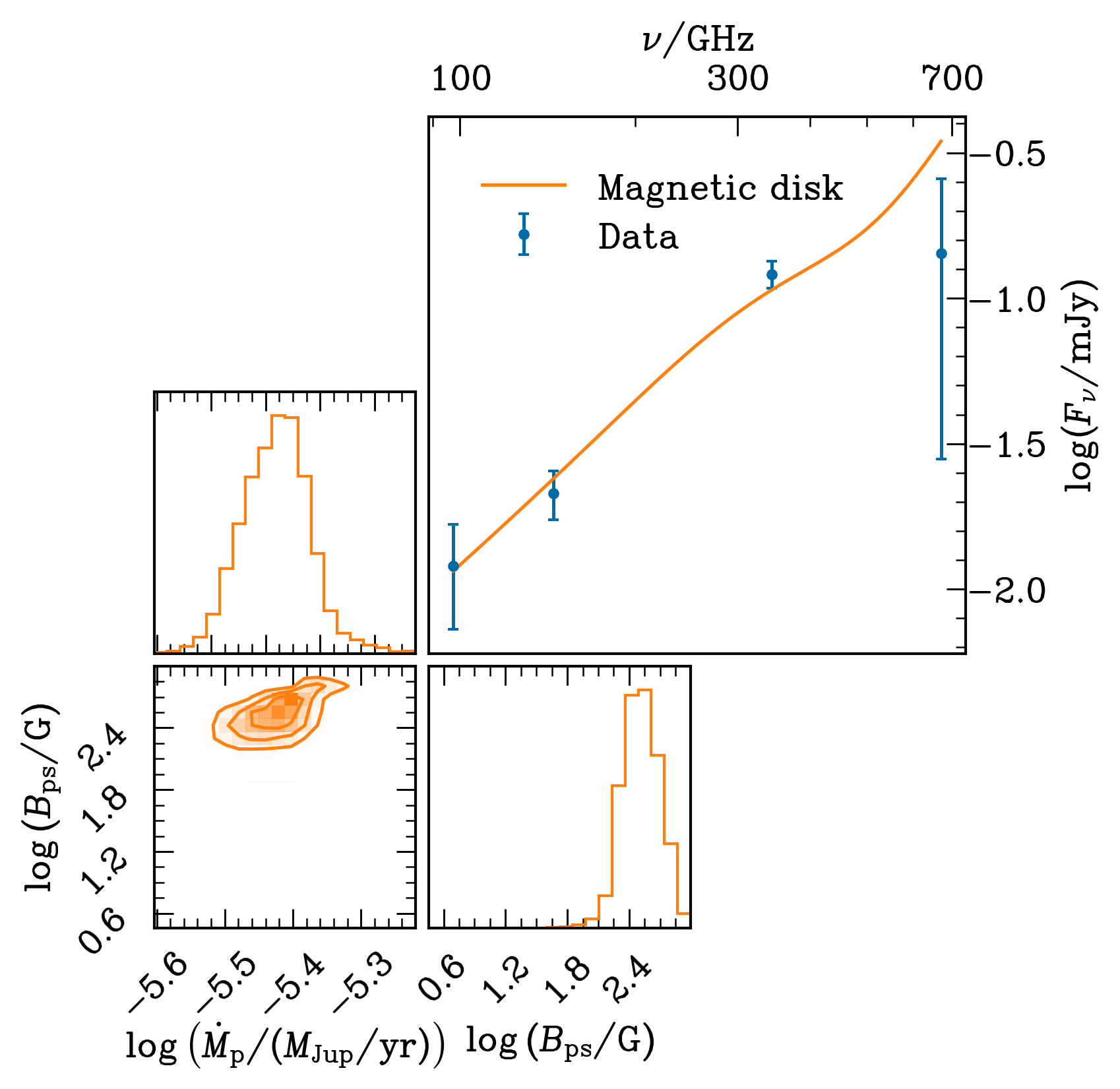}}
  \caption{Results using the magnetic disc model described in Sec.\,\ref{sec:mhd_disk}. This takes into account the free-free from metals and ${\rm H}^{-}$ and the bound-free from ${\rm H}^{-}$ described in Sec.\,\ref{sec:H_minus}. On the upper right, we have the best fit for the data of Bands\,3, 4, 7, and 9 represented with their error bars using the fluxes of Table~\ref{tab:PDS70c}. The modelled flux is represented by the solid line, obtaining a partially thick free-free emission with a spectral index of 1.76. At the bottom left, we have the corner plot with the best fit parameters $\log \left({\dot{M}_{\rm p}}/({M_{\rm Jup}/\rm yr})\right)=-5.37^{+0.06}_{-0.12}$ and $B_{\rm ps}=610^{+98}_{-375} \rm G$ at 1\,$\sigma$. The minimum reduced $\chi^2$ is 2.37.}
  \label{fig:mhd_disk}%
\end{figure}

For the magnetic disc, we chose uniform priors for $\log(B_{\rm ps}/ \rm G)$ between -4 and 4 and $\log\left(\dot{M}_{\rm p}/({M_{\rm Jup}/\rm yr})\right)$ between -10 and -2. We also added a penalty in the log likelihood for the magnetospheric accretion criterion by subtracting the quantity $x=10 \times \frac{|R_{\rm T}-R(T_{\rm c}=1000 \, \rm K)|}{R(T_{\rm c}=1000 \, \rm K)}$ for the cases in which $R_{\rm T}<R(T_{\rm c}=1000 \, \rm K)$. The resulting best fit is shown in Fig.~\ref{fig:mhd_disk}. Now we could produce spectral indices of $(\alpha_{97.5}^{145},\alpha_{145}^{343.5}, \alpha_{343.5}^{671}) = (1.85, 1.71, 1.72)$ that could marginally fit the data, and overestimate Band 9 with a flux $F_{671}$ of 339 $\mu {\rm Jy}$.

\begin{figure}
  \centering
  \resizebox{0.8\hsize}{!}{\includegraphics{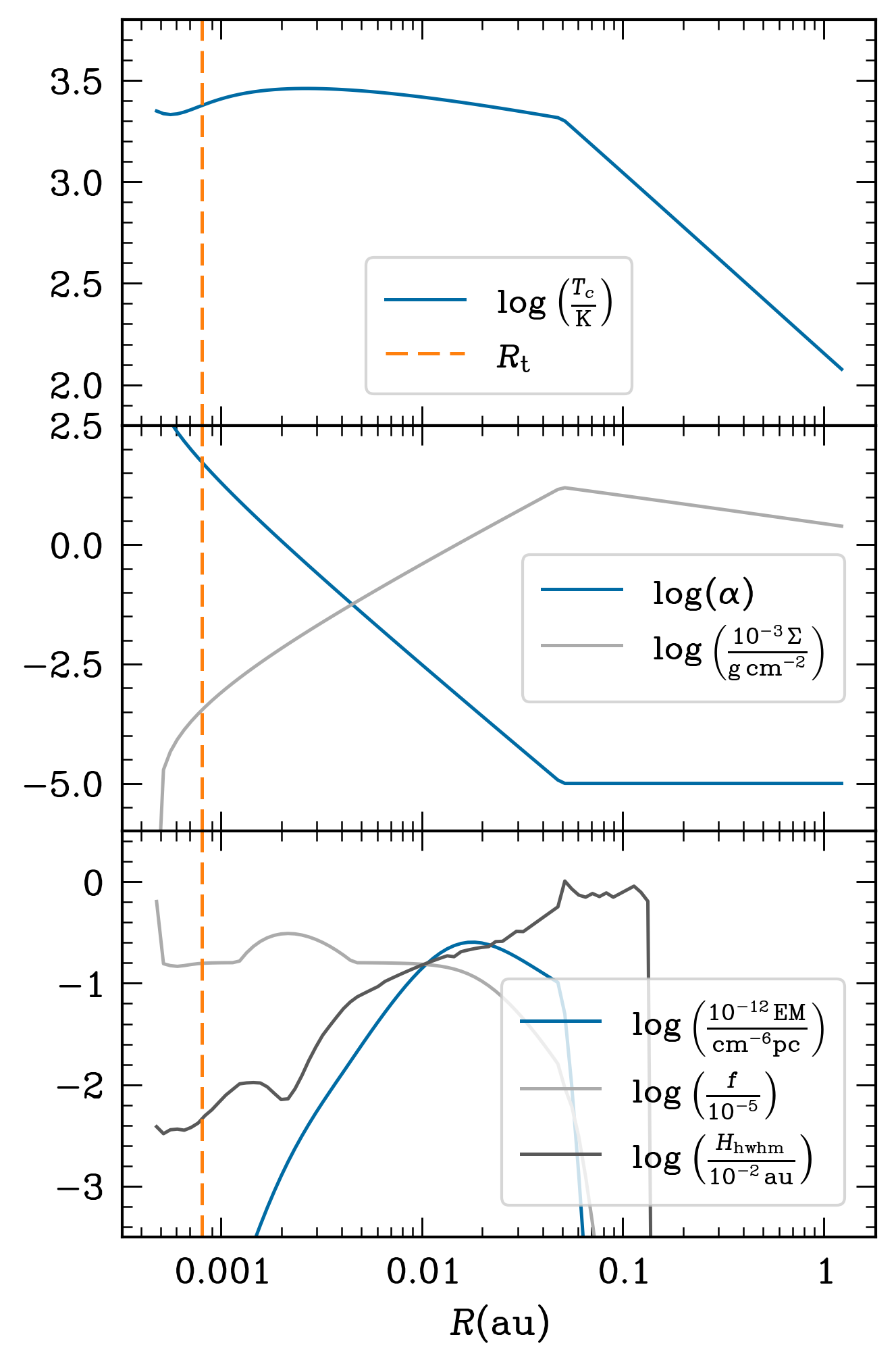}}
  \caption{Physical conditions including metal free-free emission in the best-fit MHD model. $\alpha$ is the viscosity parameter, $T_{\rm c}$ is the mid-plane temperature calculated according to Eq.~\ref{eq:Tc}, $\Sigma(R)$ is the surface density, EM is the emission measure, $f$ is the ionisation fraction in the mid-plane, and $H_{\rm hwhm}$ is the height at which the density of free electrons is half the density of free electrons in the mid-plane (half-width at half-maximum). The vertical dashed red line is the truncation radius, $R_{\rm T}=1.69 R_{\rm Jup}$, and the disc total mass is $1.93 M_{\rm Jup}$.}
  \label{fig:mhd_profiles}%
\end{figure}

From Fig.~\ref{fig:mhd_profiles} we can see that the model satisfies the magnetospheric accretion criterion ($R_{\rm T}>R(T_{\rm c}=1000 \, \rm K)$). The radius of the peak EM, with   $\rm EM \sim 10^{11.5}\, cm^{-6}\, pc$, is   $R\sim 0.02 \, {\rm au}$, and both EM and $R$ approximate those required by the uniform slab model.  The size of the magnetic disc model is smaller than the $1/10$ the Hill radius, which is reminiscent of the upper limit to the extension of the sub-millimetre signal observed in SR\,12c \citep[][]{Wu2022ApJ...930L...3W}.

Because of the residual ionisation stemming from collisional ionisation of metals, part of the atomic phase of the CPD can be seen as an atomic plasma. This state of matter is characterised by temperatures of $T \sim 1000 - 2000$\,K. The free electrons released by the metals interact with both neutral hydrogen and ionised metals, and thus produce the radio signal.

\begin{figure}
  \centering
  \resizebox{0.8\hsize}{!}{\includegraphics{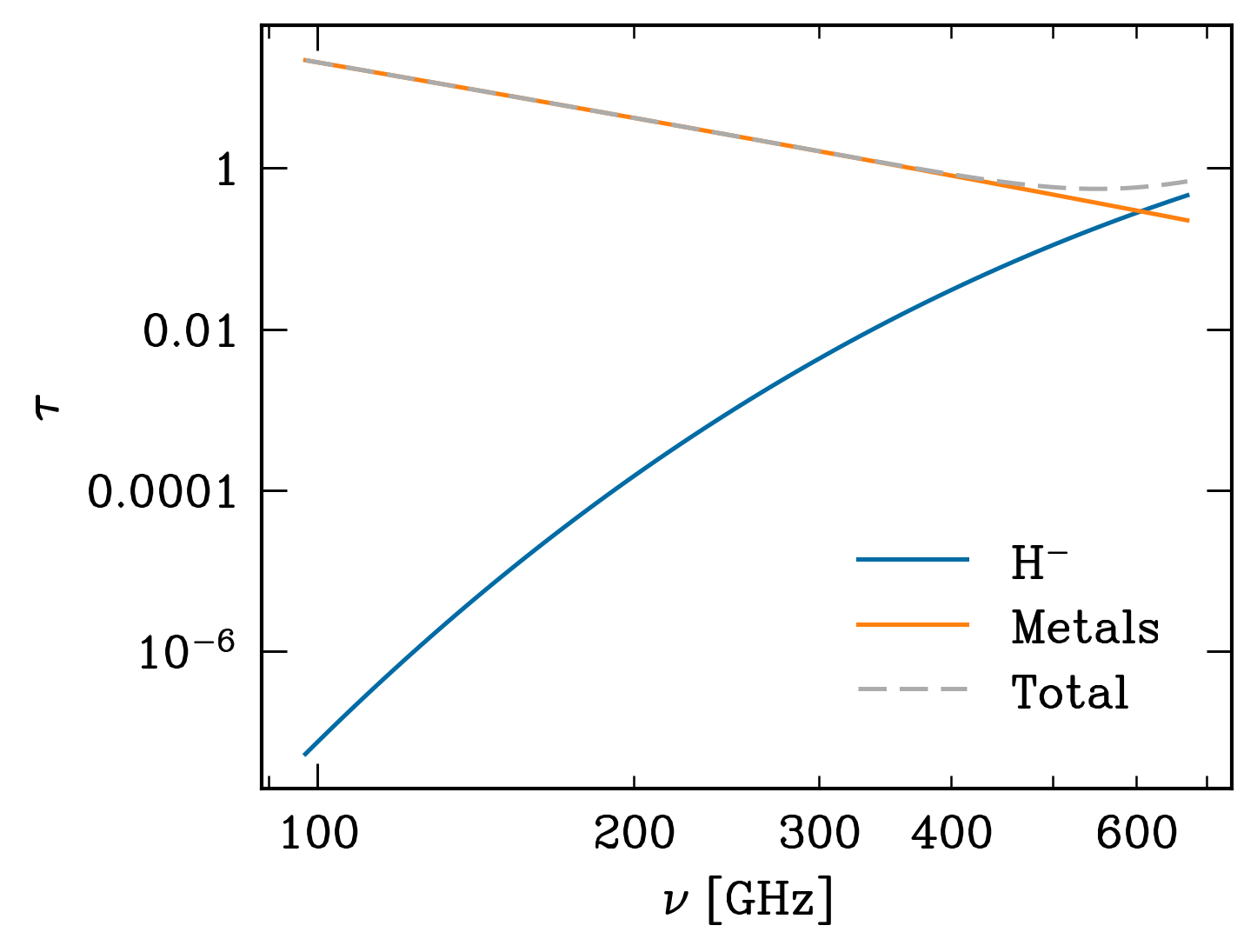}}
  \caption{Optical depth of ${\rm H}^{-}$ (including free-free and bound-free) and metals, in the annulus that makes the greatest contribution to the flux for the magnetic disc model (Sec.~\ref{sec:mhd_disk}). The solid blue and orange lines correspond to the optical depth by  ${\rm H}^{-}$ and metals, respectively. The dashed line is the total optical depth.}
  \label{fig:optical_depth}%
\end{figure}


Atomic plasma radiation in the magnetic disc model can marginally fit the data. However, an optically thick spectrum persists at higher frequencies and the drop in Band\,9 cannot be reproduced. This is because, even though the magnetic disc reaches the necessary EMs for free-free emission from ionised metals, ${\rm H}^{-}$ free-free also contributes to the total optical depth. In Fig.~\ref{fig:optical_depth} the optical depth of metals behaves exactly as we would like it to, going from optically thick to thin from 97.5\,GHz to $\sim$500\,GHz. However, the optical depth of ${\rm H}^{-}$ begins to dominate above 500\,GHz, and the total optical depth remains $\tau \sim 1$ without reaching the optically thin regime.

The atomic plasma model, however, depends heavily on the accuracy of the extrapolation from the  ${\rm H}^{-}$ opacities, specially the free-free opacity that dominates in the sub-millimetre continuum. \citet{Gray2022oasp.book.....G} was inspired by works that calculated opacities of up to $10\,\mu$m and in a range of temperatures from 2500 to 10000\,K. This is very important here, since we are calculating the opacity at millimetre wavelengths, and in a cold disc where the majority of the flux comes from temperatures of $T\approx1000$--2000\,K.



\section{Discussion}
\label{sec:discussion}

\subsection{Which models fit the data} \label{sec:discussion_model}

The best fits obtained for the dusty disc model, shown in Fig.\ref{fig:bestfitdust}, have spectral indices of $\alpha >2.35$ and do not fit the data, even between Bands 3 and 7. Adding a jet, either optically thin or thick, does not change this result. However, we observe that the bulk of the flux stems from the outer regions of the disc. If the temperature is low enough ($T_{\rm floor}<15$), then a dusty disc can fit the data, as is shown in Fig.~\ref{fig:bestfitzhu_T}. This can be understood from the Wien displacement law, which states that at such low temperatures the turnover of the black body spectrum is close to the Band\,9 frequency. However, these low temperatures are unlikely, as is mentioned in Sec.\,\ref{sec:dust_comp}. Band\,9 shows a lower temperature limit of 21\,K at 34\,au, and the presence of CO in the PDS\,70 cavity suggests temperatures of at least 30\,K \citep{Law2024ApJ...964..190L}. Therefore, we conclude that it is not possible to fit the observed SED using only a dusty disc model with the current derivations of $T_{\rm floor}$.

If viscous heating is efficient and reaches H-ionising temperatures, our best fit with H\,{\sc i} free-free from the H\,{\sc ii} disc reaches $\rm EM \sim 10^{26-31} \, {\rm cm}^{-6}\, {\rm pc}$ (see Figs.~\ref{fig:profilesfixedalpha} and \ref{fig:profilesfixedMpdot}), which seems overwhelming and would approximate the radio spectrum of an early-type star. This results in an optically thick SED which overshoots the flux density in Band\,9.

A uniform slab model can inform one about the parameters needed to produce the Band\,9 drop, with an EM of $\sim 10^{11-12} \, {\rm cm}^{-6}\, {\rm pc}$, temperatures of $T\sim 10^3-10^4\,{\rm K}$ and sizes of $\sim$0.1\,au. Comparing these EMs and temperatures with those from the H\,{\sc ii} disc (see the implicit dependence of ${\rm EM}$  and $T$ in Figs.~\ref{fig:profilesfixedalpha} and \ref{fig:profilesfixedMpdot}), we see that  EMs in an H\,{\sc ii} CPD would require low ionisation fractions, and hence  fine-tuning of the temperature. Since $f(T)$ is extremely sensitive to $T$, we conclude that the CPD is probably not hot enough to ionise  hydrogen,
as, unless the temperature field is  `just right', hydrogen-ionisation results in huge ${\rm EM}$ and  a   very optically thick SED  (and inconsistent, at 2.6\,$\sigma$,  with the non-detection in Band\,9).

Finally, if the planet has a significant magnetic field ($B_{\rm ps} > 200 \,{\rm G}$), magnetospheric accretion occurs, the temperature drops to $\sim 10^3$\,K, and the signal stems from free-free emission from ionised metals at lower frequencies ($\nu \lesssim 500$\,GHz) and from ${\rm H}^{-}$ at higher frequencies ($\nu>500$\,GHz). This atomic plasma radiation model maintains a partially optically thick SED above 500\,GHz and also overshoots Band\,9.

\subsection{Surface shock on the CPD}
\label{sec:CPD_shock}

The small EMs from the uniform-slab models are reminiscent of shock models, as was studied by \citet{Aoyama2018ApJ...866...84A} and \citet{Aoyama2020arXiv201106608A} for the CPD surface shock and  \citet{Aoyama2020arXiv201106608A} for the photospheric shock.  These shocks arise from the supersonic, nearly vertical gas flows that collide with the CPD or planetary surfaces, as was predicted in detailed hydro-dynamical simulations \citet{Tanigawa2012ApJ...747...47T, Szulagyi2020ApJ...902..126S}. The strong shock heats the gas and results in a thin, non-equilibrium layer of ionised hydrogen.

Because of the high temperatures ($T \gtrsim 10^4\,{\rm K}$) in the post-shock gas, we can neglect contributions from ${\rm H}^{-}$ to the net opacity. The bulk of the mass is at a relative constant temperature of $\sim 10^4\,{\rm K}$. Since the most emission originates from the region with $T \sim 10^4$\,K, we refer to $10^4$\,K as the effective temperature for free-free emission \citep{Aoyama2018ApJ...866...84A,Aoyama2020arXiv201106608A}.  As is seen in the uniform slab model of Sec.~\ref{sec:uniform_slab} for temperatures of $10^4$\,K, an effective radius of $\sim 52^{+5}_{-5} \,{R_{\rm Jup}}$ is required to fit the SED. This indicates that the flux in the radio should come from shocks on the CPD and neither from the photosphere nor accretion funnels  (whose sizes are $\sim 2\,{R_{\rm Jup}}$).

To firmly rule out photospheric shocks and accretion funnels, we estimated the temperature and EM needed for a fixed size of $R_{\rm max} = 2\,{R_{\rm Jup}}$ in the uniform slab model. The best fit parameters are $\log\left( T/\rm K \right)=6.83^{+0.04}_{-0.10}$ and $\log\left(\rm EM/(cm^{-6}\,pc) \right) = 15.75^{+10.11}_{-0.14}$. Such  temperatures are  only achieved in the post-shock region of an object $>200\,M_{\rm Jup}$ with a $2\,R_{\rm Jup}$ radius, which is outside of the planetary regime. High temperatures, in excess of $\sim 10^6$\,K,  also quench H$\alpha$ radiation as recombination becomes less probable. Therefore a photospheric shock, or free-free emission from the accretion funnels, cannot be  the source of the radio signal  as such shocks would require excessively high temperatures, given the free-fall velocities in the planetary regime. Note that  part of the H$\alpha$ signal from  PDS\,70c could still stem from a $T \sim 10^4$\,K shock on the planetary surface, even if such a shock does not account for the radio signal.

\begin{figure}
  \centering
  \resizebox{\hsize}{!}{\includegraphics{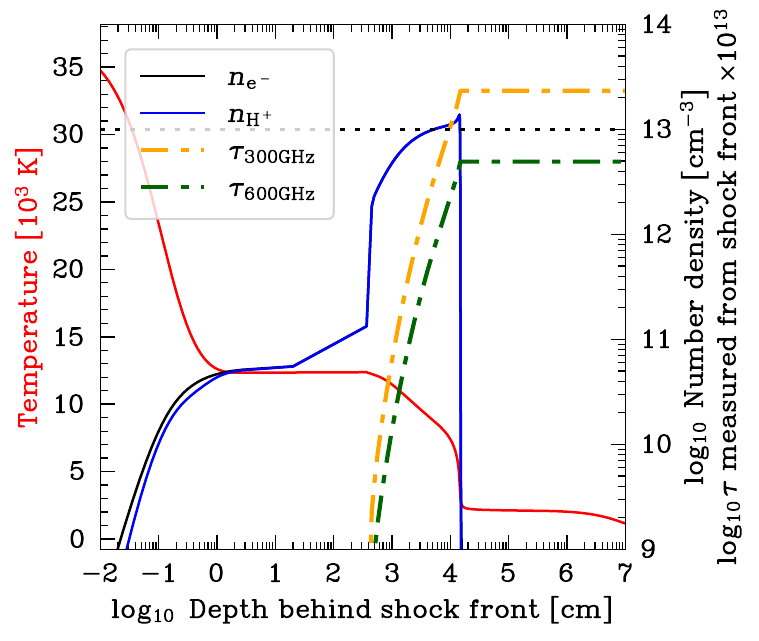}}
  \caption{Physical conditions of the post-shock region for $\varv_{0}=30\, \rm km\,s^{-1}$ and $n_{0}=10^{15}{\rm cc}^{-1}$. The temperature, free electron local density ($n_{\rm e^{-}}$), and H\,{\sc ii} local density ($n_{\rm H^{+}}$) are the solid red, black, and blue lines, respectively. The optical depths at 300 and 600\,GHz ($\tau_{300\, \rm GHz}\,\rm and\,\tau_{600\, \rm GHz}$) are the dashed green and yellow lines, where the dotted black line corresponds to $\tau=1$.}
  \label{fig:aoyama30}%
\end{figure}
\begin{figure}
  \centering
  \resizebox{\hsize}{!}{\includegraphics{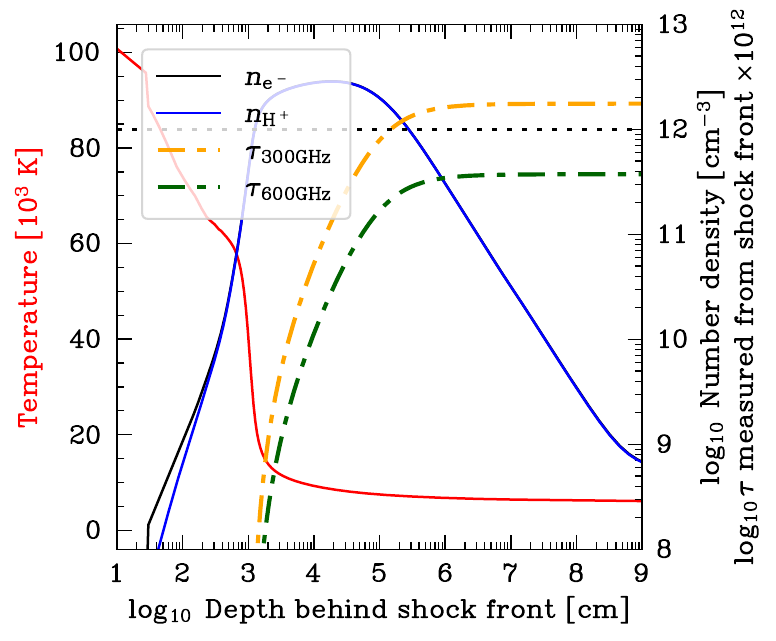}}
  \caption{Physical conditions of the post-shock region for $v_{0}=50\, \rm km\,s^{-1}$ and $n_{0}=10^{12}{\rm cc}^{-1}$. Annotations follow from Fig.~\ref{fig:aoyama30}.}
  \label{fig:aoyama50}%
\end{figure}

To confirm that the post-shock emission on the CPD surface can indeed reproduce the observed radio emission, we performed two simulations of shock-heated gas following \citet{Aoyama2018ApJ...866...84A}, whose results are shown in Figs.~\ref{fig:aoyama30} and \ref{fig:aoyama50}. These correspond to pre-shock velocities and neutral hydrogen number densities of $(\varv_0\, , n_0)=(30\,{\rm km\,s^{-1}}\, , 10^{15}\, \rm cm^{-3}$) and  $(50\,{\rm km\,s^{-1}}\, , 10^{12}\, \rm cm^{-3}$) respectively. In both cases, the emission at 300\,GHz is optically thick ($\tau>1$) but turns over into an optically thin spectrum at 600\,GHz  ($\tau \sim$\,a few times\,$0.1$). This is also the case for the SED of PDS\,70c obtained with the uniform slab model, with a best fit of  $\tau_{300\,\rm GHz}\approx 3$ and $\tau_{600 \rm GHz}\approx 0.6$.

In these simulations, the optical-depth weighted temperatures are $\sim 7000$ and $\sim 7500$\,K, for $\varv_0=30\,\rm km\,s^{-1}$ and $\varv_0=50\,\rm km\,s^{-1}$, respectively. A uniform slab at $7000$\,K requires a radius of $R_{\rm max}=62\, R_{\rm Jup}$, which is in between the values obtained for the uniform slab  models at $10^3$\,K and $10^4$\,K shown in Fig\,\ref{fig:pill}. We can estimate the planetary mass required to achieve a pre-shock velocity of $\varv_0>30\,{\rm km\, s^{-1}}$ at $62\,R_{\rm Jup}$, by assuming that $\varv_0$ is the free-fall velocity,
\begin{equation}
  \label{eq:velocity}
  \varv_0=\sqrt{\frac{2 G M_{\rm p}}{r}}.
\end{equation}
A minimum planetary  mass of 15.7\,$M_{\rm Jup}$ is required, which is 2.1\,$M_{\rm Jup}$ above the $2\sigma$ upper limit of 13.6\,$M_{\rm Jup}$ estimated for the dynamical mass of PDS\,70c \citep{Trevascus2025A&A...698A..19T}. This minimum planetary mass could  be reduced by considering the increasing $\varv_0$ at smaller radii, and its resulting  radio flux, but this is beyond the scope of this work.

Using a mean molecular weight of $\mu=1.35$, $\varv_0=30\,{\rm km s^{-1}}$ and $n_0=10^{15} \,{\rm cc^{-1}}$ implies a mass flux of $\mu m_{\rm H} \times n_{0} \times \varv_{0} = 6.75 \times 10^{-3}\,{\rm cm^{-2}\, s^{-1} \,g}$. If this flux is is spatially constant and falls on both sides of a circular slab with a radius of $62\,R_{\rm Jup}$, the mass influx rate within this area is $0.1\, M_{\rm Jup}\,{\rm yr}^{-1}$. For $\varv_0=50\,{\rm km s^{-1}}$ and $n_0=10^{12} \,{\rm cc^{-1}}$, the crude estimate of the mass influx rate is instead $2\times10^{-4}\, M_{\rm Jup}\,{\rm yr}^{-1}$, for simplicity using the same effective radius of $62\,R_{\rm Jup}$. These rough estimates of the accretion rate may seem extremely high. However, not all of the mass falling on the CPD is necessarily accreted by the planet because of equatorial decretion and recycling of the CPD mass along the planetary wakes \citep{szulagyi2016MNRAS.460.2853S, Batygin2018AJ....155..178B}.

Moreover, while we assumed a constant mass flux across the entire CPD, as in \citet{Aoyama2018ApJ...866...84A}, a concentration of mass flux towards the inner region could significantly lower the mass-accretion-rate estimate, while preserving the radio flux. 
The larger  $\varv_0$ and  radio intensity at smaller radii could produce the same radio flux with a smaller mass accretion rate. 
The radial distribution of  CPD mass-loading is, however, an open question, with some studies favouring the outer regions of CPDs  \citep[][]{Tanigawa2012ApJ...747...47T,Marleau2024ApJ...964...70M}.

If the observed radio signal from PDS\,70c is entirely due to H\,{\sc i} free-free emission from the CPD surface shock, the emission from the CPD itself must   be faint. This places an upper limit on the planetary  accretion rate. Using the magnetic disc model from Sec.\,\ref{sec:mhd_disk} with the best fit value of $B_{\rm ps}=610$\,G, we obtain an upper limit of $\log \left(\dot{M}_{\rm p}/(M_{\rm Jup}/\rm yr)\right)<-5.42$ using the 3\,$\sigma$ flux limit in Band\,7. In the case of a smaller magnetic field (for example $B_{\rm ps}=200$\,G), the upper limit for the accretion rate diminishes to $\log \left(\dot{M}_{\rm p}/(M_{\rm Jup}/\rm yr)\right)<-5.77$.


The non-detection of PDS\,70b can be understood in the context of the CPD-surface shock if  PDS\,70b is   less massive than PDS\,70c \citep[e.g.][]{Trevascus2025A&A...698A..19T}. In this case, PDS\,70b would have the same $\varv_0$ at a smaller radii. As an illustration, for a 3\,$M_{\rm Jup}$ planet, the pre-shock velocity $\varv_0=30\,\rm km\, s^{-1}$ is obtained at $R_{\rm shock, b} = 12\,R_{\rm Jup}$. This means that the radii for the shocked region where free-free emission arises is smaller and consequently the total flux decreases. As a rough approximation, the radio flux from PDS\,70b would be lower than PDS\,70c by a factor corresponding to the ratio of the shocked surface area, or $\sim (R_{\rm max}/R_{\rm shock, b}) \sim  (62/12)^2  \sim 30$.

\section{Conclusion}
\label{sec:conclusion}

We report on new ALMA observations of PDS\,70, focussing on the point-source signal coincident with PDS\,70c, with  new detections in Bands\,4 and 7, and a marginal detection in Band\,3. The spectrum across Bands 3, 4 and 7 is consistent with optically thick emission, with $\alpha = 2.0\pm 0.2$.  However, a non-detection in Band\,9 breaks this trend, with a flux density lower than an optically thick extrapolation of lower frequencies by at least $2.6\,\sigma$.  The Band\,7 signal from PDS\,70c appears to be constant within 10\% rms, from May to December 2023. Since the multi-frequency measurements are coeval within two months, we assume that the SED collected here is constant. 

The non-detection in Band\,9 is particularly surprising, given the model predictions for bright signal at short sub-millimetre wavelengths, and suggest a very dust poor disc, whose radio signal stems from   partially thick free-free radiation.  We interpret the observed SED in an adaptation of parametric CPD models, including the free-free emission from the disc itself. 

A purely dusty disc cannot reproduce the  SED of PDS\,70c, even with the inclusion of free-free radiation from a jet. However,  the  H\,{\sc i} free-free continuum from the CPD yields an  optically thick SED, and  marginally  accounts for the data (within 3\,$\sigma$ of Band\,9). This free-free signal  is intrinsic to the CPD and a  product of the high temperatures reached by viscous heating in these analytic CPD models. Such an H\,{\sc ii} disc is at odds with the atomic or molecular environments expected in the  context of planetary accretion.

Given an optically thick free-free spectrum, any steeper component is negligible, and the dust-to-gas mass ratio $\zeta$ must be very low, $\log(\zeta)<-4.80$. Such a dust-deprived environment could be the result of strong dust filtering inside the planet-induced gap.


Uniform slab models indicate the need for lower EMs than are obtained in constant-viscosity CPDs to account for the turnover to partially thin free-free emission above Band\,7 (see Fig.\,\ref{fig:pill}). The free-free signal would then stem from an atomic plasma, with  residual ionisation from metals. The required EMs and temperatures, of $T \sim 2000$\,K, are naturally produced in a magnetic disc models where large disc viscosity (up to $\alpha \sim 0.1$) results from a strong magnetic field. 

Magnetic disc models result in similar radio SEDs as for the H\,{\sc ii} disc, but without reaching H-ionising temperatures, and with comparatively moderate EMs ($\rm EM \sim 10^{11.5}\, cm^{-6}\, pc$). However, because of the opacity of $\rm H^{-}$ free-free, an optically thick SED persists despite the lower EMs, and still overshoots the Band\,9 flux density.

The CPD surface shock model is the closest one to the uniform slab requirements, and is therefore the only viable interpretation of the observed SED, including the Band\,9 drop. Such a surface shock requires inefficient planetary accretion with a strong mass flux onto the CPD, and correspondingly strong equatorial mass loss.

The radio SED of PDS\,70c rules out a planetary surface shock, or accretion funnels, as the source of the signal. Such small emitting regions would require temperatures  in excess of $10^6\,$K to reach the observed flux densities, which cannot be produced in shock models in the planetary mass regime.

\begin{acknowledgements}
  We thank the referee for constructive comments.   O.D., S.C.,  M.C. and P.W. acknowledge support from Agencia Nacional de Investigaci\'on y Desarrollo de Chile (ANID) given by FONDECYT Regular grants 1211496, ANID MAGISTER NACIONAL BECAS CHILE/2025-22250525, ANID PFCHA/DOCTORADO BECAS CHILE/2018-72190574, ANID project Data Observatory Foundation DO210001, FONDECYT postdoctoral grant 3220399 and Millennium Science Initiative Program Center Code NCN2024\_001, FONDECYT grant 3220399 and ANID -- Millennium Science Initiative Program -- Center Code NCN2024\_001. The work of OC was supported by the Czech Science Foundation (grant 21-23067M), the Charles University Research Centre program (No. UNCE/24/SCI/005), and the Ministry of Education, Youth and Sports of the Czech Republic through the e-INFRA CZ (ID:90254). This paper makes use of the following ALMA data:
{\tt ADS/JAO.ALMA\#2022.1.00592.S},
{\tt \#2022.1.01477.S},
{\tt 2022.1.00893.S}.
ALMA is a partnership of ESO (representing its member states), NSF (USA) and NINS (Japan),
together with NRC (Canada), MOST and ASIAA (Taiwan), and KASI
(Republic of Korea), in cooperation with the Republic of Chile. The
Joint ALMA Observatory is operated by ESO, AUI/NRAO and NAOJ.

\end{acknowledgements}

\bibliographystyle{aa} 

\bibliography{refs} 


\end{document}